\documentclass[aps,preprint,letterpaper,amsmath,amssymb]{revtex4}
\usepackage{graphicx}
\usepackage{dcolumn}
\usepackage{bm}
\begin{document}

\newcommand{\rf}[1]{(\ref{#1})}

\newcommand{\bfomega}{ \mbox{\boldmath{$\omega$}}}

\title{
Origins of fractality in the growth of complex networks}

\author{Chaoming Song$^1$, Shlomo Havlin$^2$, and Hern\'an A. Makse$^1$}

\affiliation{ $^1$ Levich Institute and Physics Department,
City College of New York, New York, NY 10031, US\\
$^2$ Minerva Center and Department of Physics, Bar-Ilan
University, Ramat Gan 52900, Israel}

\date{\today}

\begin{abstract}

{\bf Complex networks from such different fields as biology, 
technology or sociology share similar organization principles. 
The possibility of a unique growth mechanism for a variety of
complex networks in different fields such as biology, technology
or sociology is of interest, as it promises to uncover the
universal origins of collective behavior. The emergence of
self-similarity in complex networks raises the fundamental
question of the growth process according to which these structures
evolve. Here, we use the concept of renormalization as a mechanism
for the growth of fractal and non-fractal modular networks. We
show that the key principle that gives rise to the fractal
architecture of networks is a strong effective ``repulsion"
(disassortativity) between the most connected nodes (hubs) on all
length scales, rendering them very dispersed. More importantly, we
show that a robust network comprised of functional modules, such
as a cellular network, necessitates a fractal topology, suggestive
of an evolutionary drive for their existence.}

\end{abstract}

\maketitle

An important result in statistical physics was the generation of
fractal geometries by Mandelbrot \cite{mandelbrot,vicsek}, the
structures of which look the same on all length scales. Their
importance stem from the fact that these structures were
recognized in numerous examples in Nature, from snowflakes and
trees to phase transitions in critical phenomena
\cite{vicsek,stanley}. While these fascinating patterns are only
geometric, new forms of topological fractality have been observed
in complex networks \cite{shm} where the links rely on
interactions between the participants \cite{ab-review,vespignani}.
Examples of topological fractal networks include the hyperlinks in
the WWW, physical interactions in protein interaction networks or
biochemical reactions in metabolism \cite{shm,strogatz}. Other
complex networks such as the Internet do not share the topological
fractal property.

These fractal complex networks are characterized by the
small-world property (as given by the logarithmic dependence of
the average distance with the number of nodes) brought about by
the ``short-cuts'' in the network \cite{watts}, a very wide
(power-law or "scale-free" \cite{ba}) distribution of connections,
and a modular hierarchical structure
\cite{hopfield,ravasz,newman3,nested}. However, the fractal sets
of Mandelbrot do not exhibit these features. While in our previous
work \cite{shm}, we discovered the fractal nature of organization
in many real networks the question remained how these networks
have evolved in time. We therefore  launch a study of growth
mechanisms to understand the simultaneous emergence of fractality,
modularity, as well as the small world effect, and the scale-free
property in real world complex networks. Our results have
important evolutionary implications. They highlight an
evolutionary drive towards fractality, inspired by an increase in
network robustness. Thus, a robust modular network requires a
fractal topology. Furthermore, our analysis indicates that the
fractal clusters can be identified with the functional modules in
the case of the metabolic network of {\it E. coli}.

The ``democratic'' rule of the seminal Erd\"{o}s-R\'enyi model
\cite{erdos} (where the nodes in the network are connected at
random) was first invoked to explain the small world effect. It
was then replaced by the ``rich-get-richer'' principle of
preferential attachment \cite{ba} to explain the scale-free
property; a discovery carrying important implications on network
vulnerability \cite{barabasi-attack,cohen2000}. However these
rules  do not capture the fractal topologies found in diverse
complex networks. We find that models of scale-free networks are
not fractals (see Supplementary Materials Section \ref{modeling}).
Here, we demonstrate a new view of network dynamics where the
growth takes place {\it multiplicatively} in a correlated
self-similar modular fashion, in contrast to the uncorrelated
growth of models of preferential attachment \cite{ab-review,ba}.

We formalize these ideas by borrowing the concept of
``length-scale renormalization'' from critical phenomena
\cite{stanley}. In this paper, we will show that the emergence of
self-similar fractal networks, such as cellular ones,  is due to
the strong repulsion (disassortativity \cite{newman2}) between the
hubs at all length scales. In other words, the hubs prefer to grow
by connections to less-connected nodes rather than to other hubs,
an effect that can be viewed as an effective hub repulsion. In
this new paradigm, the ``rich" still get richer, although at the
expense of the ``poor''. In other words, the hubs grow by
preferentially linking with less-connected nodes to generate a
more robust fractal topology. In contrast, weakly anticorrelated
or uncorrelated growth leads to non-fractal topologies such as the
Internet.

\begin{figure*}
\centerline{ (a) \resizebox{8cm}{!} {
 \includegraphics{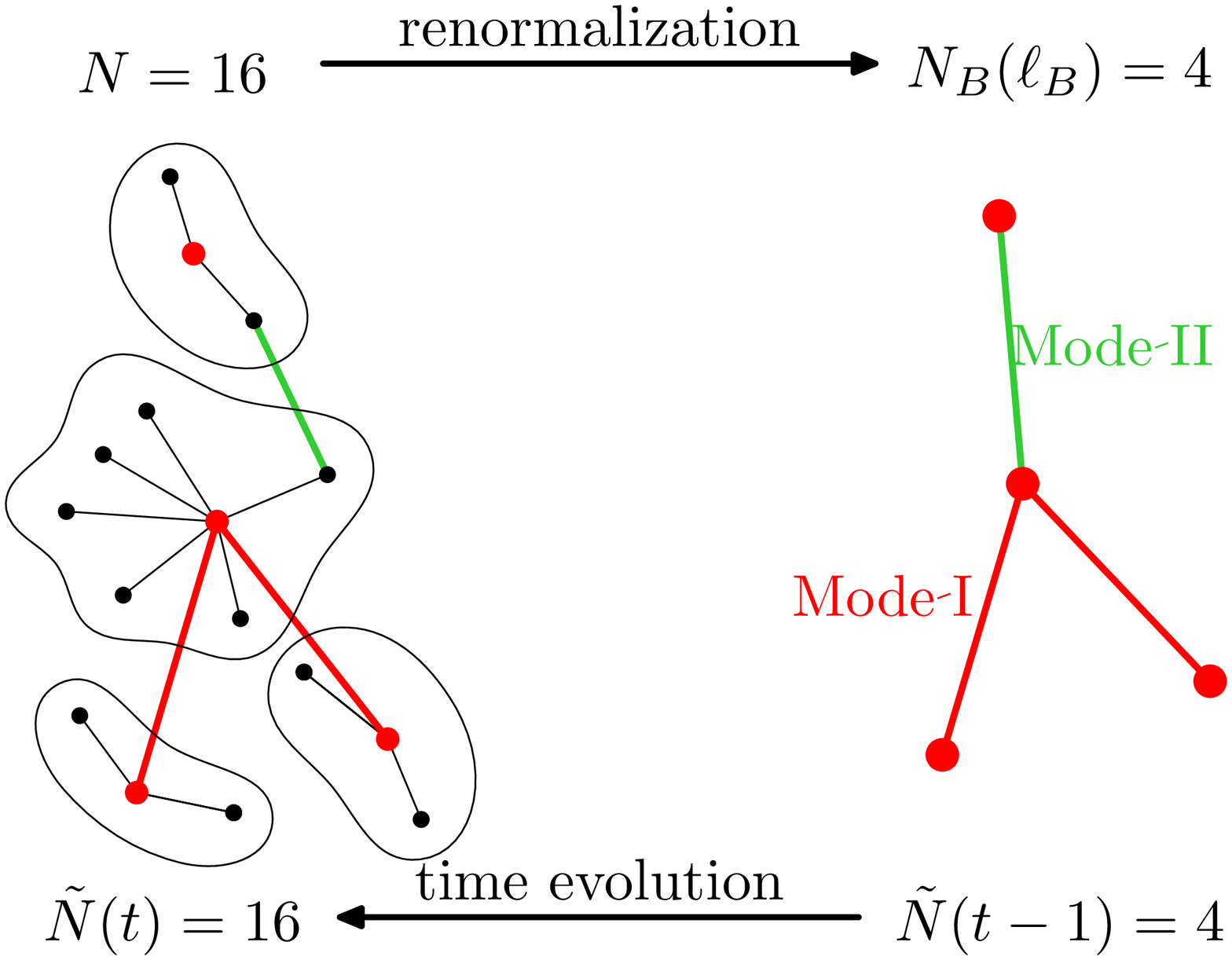}}
} \vspace{1cm} \centerline{ (b) \resizebox{8cm}{!} {
\includegraphics{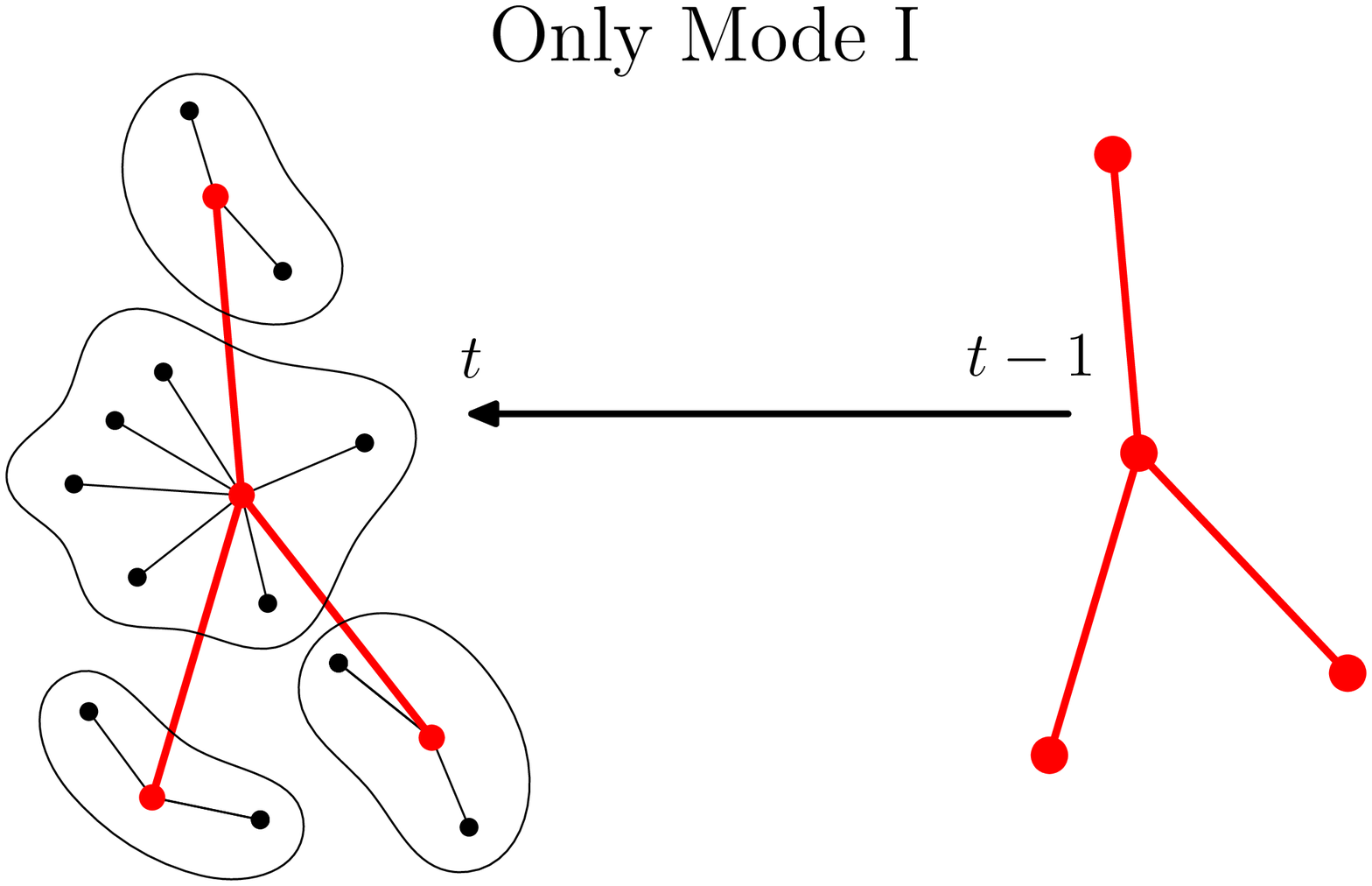}} } \vspace{1cm} \centerline{
(c)\resizebox{8cm}{!} { \includegraphics{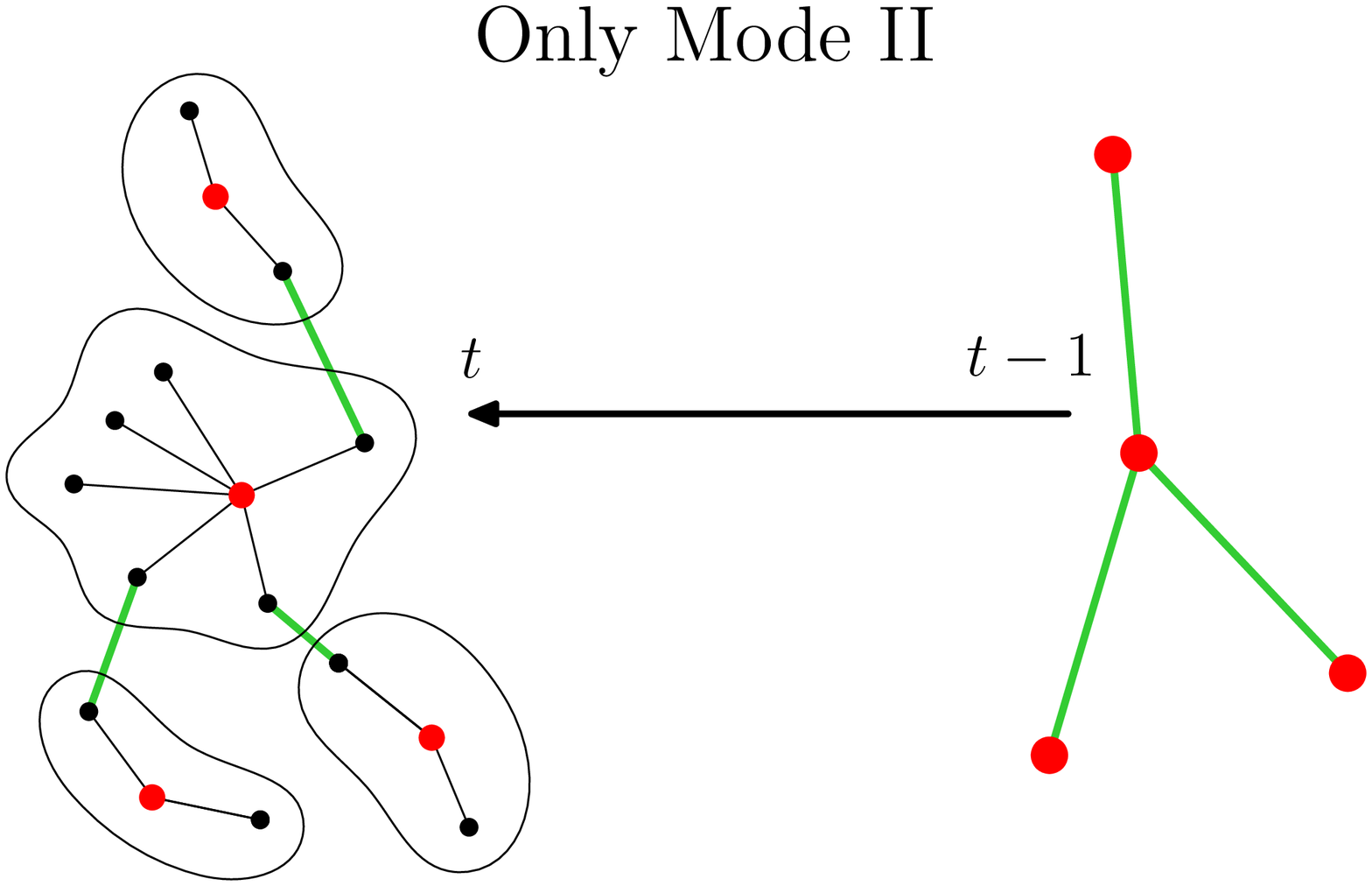}} } \caption {}
\label{renormalization}
\end{figure*}

\begin{figure*}
\centerline{ \hbox { (a)\resizebox{6cm}{!} {
\includegraphics{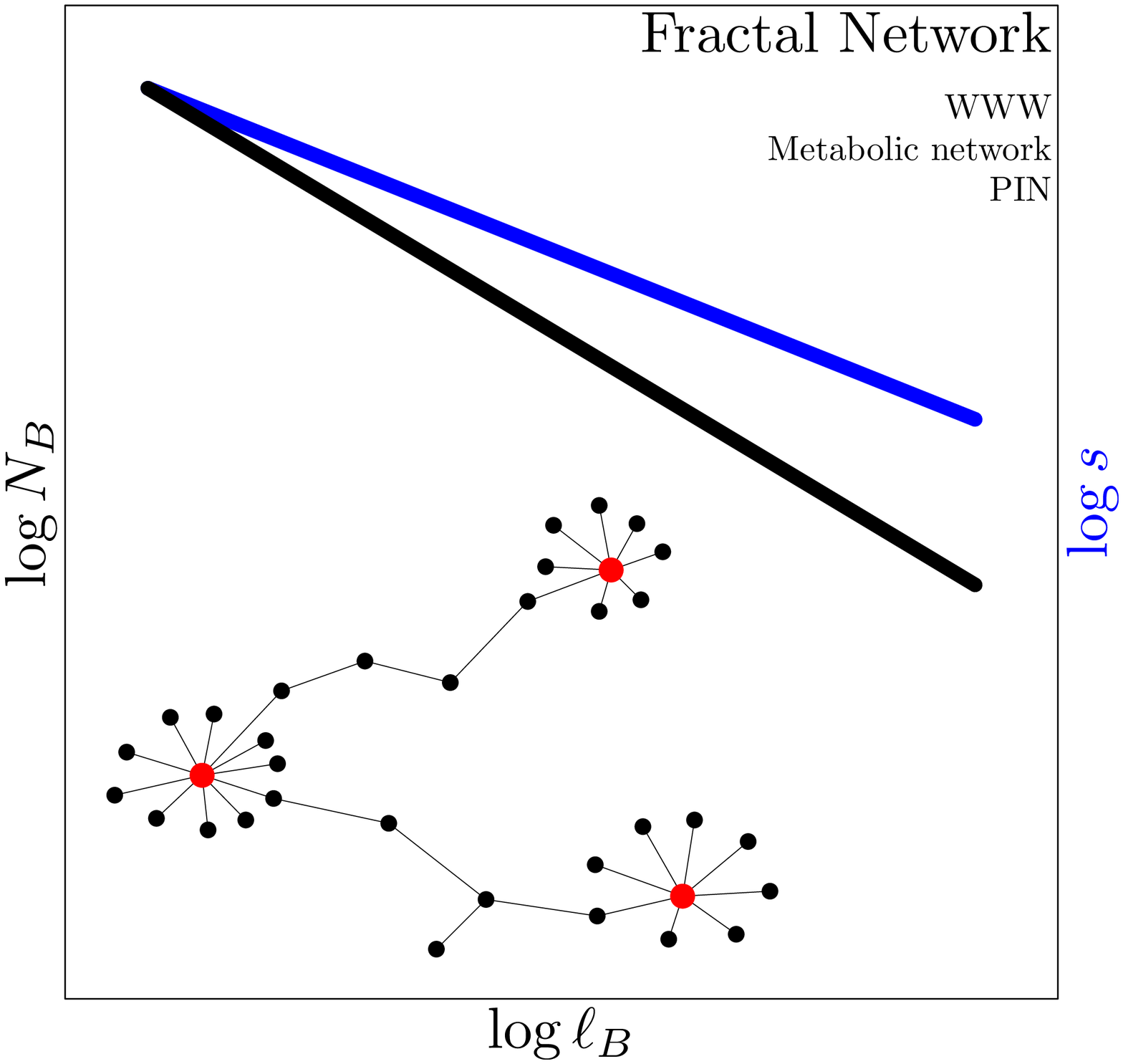}} (b)\resizebox{6cm}{!} {
\includegraphics{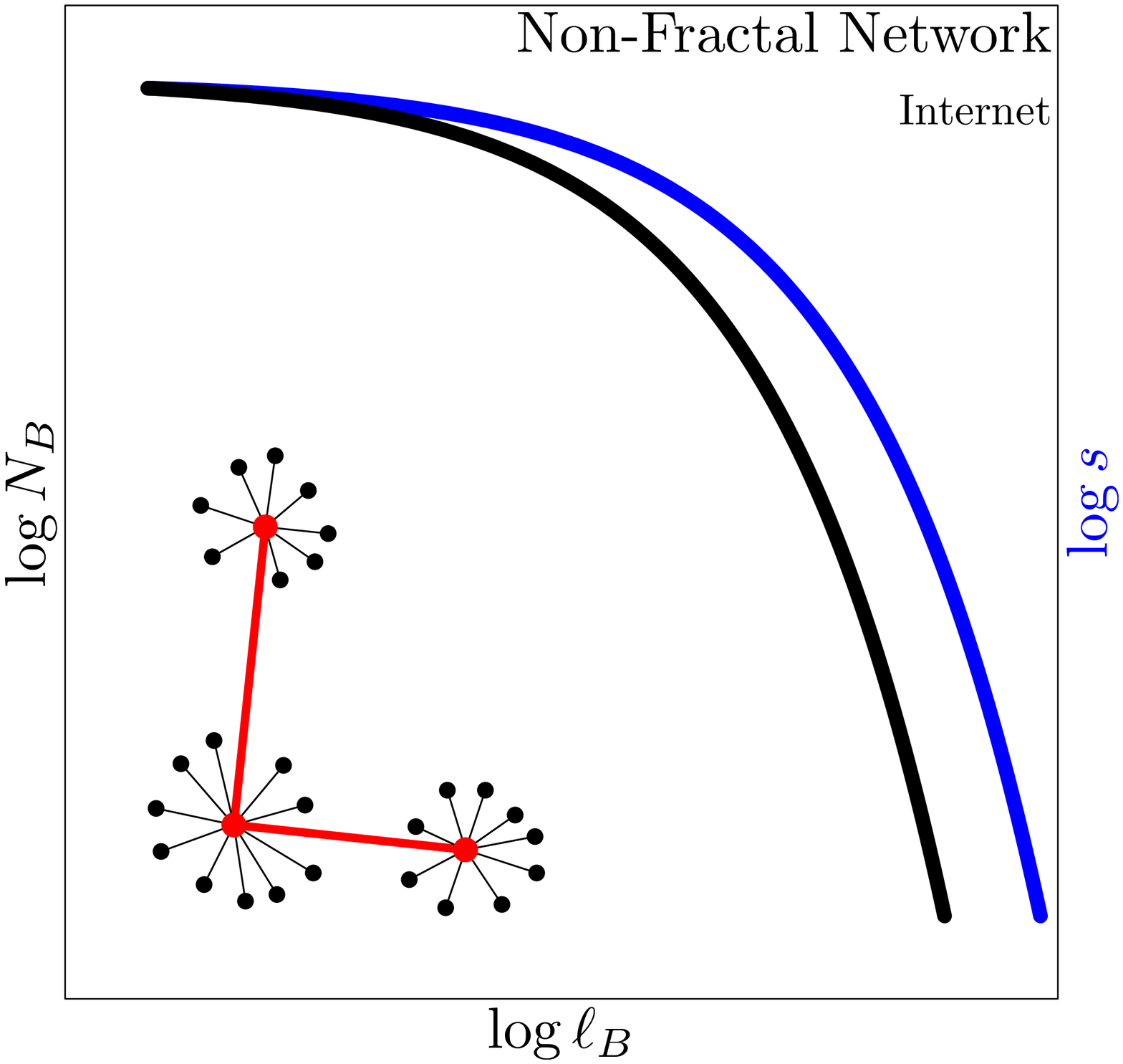}} }} \centerline{ \hbox {
(c)\resizebox{6cm}{!} { \includegraphics{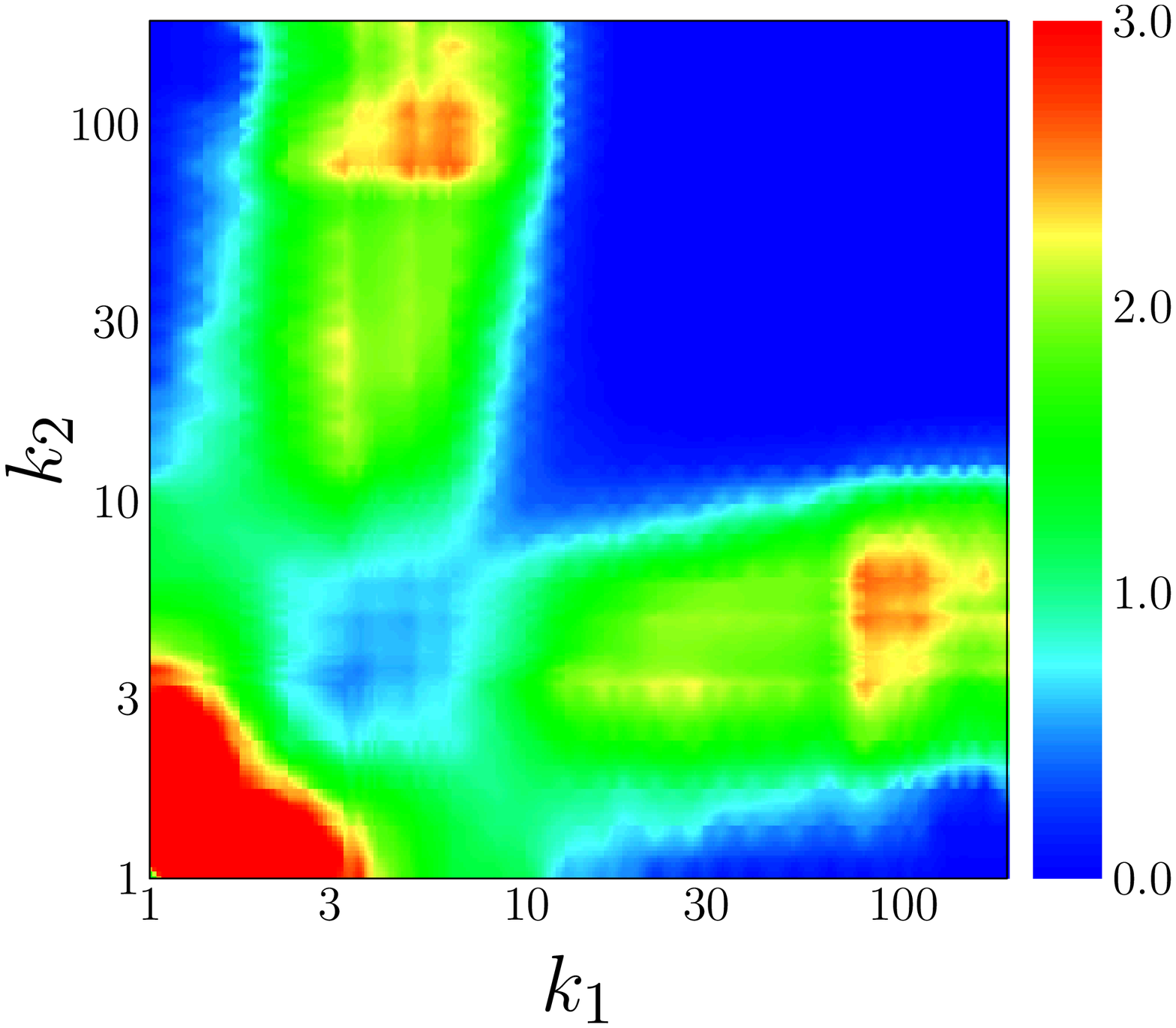}}
(d)\resizebox{6cm}{!} { \includegraphics{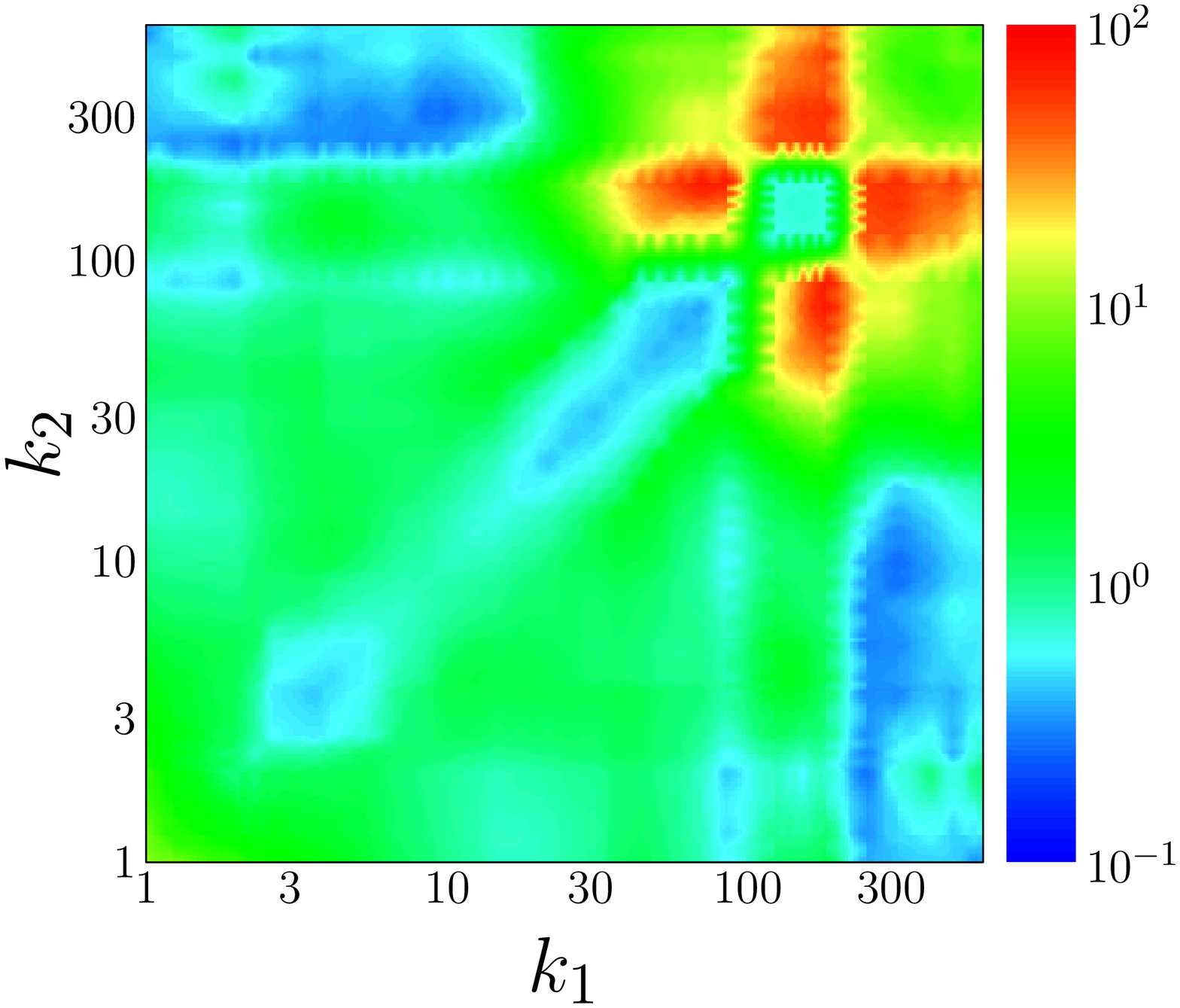}} } }
\centerline{ {(e)\resizebox{8cm}{!} {
\includegraphics{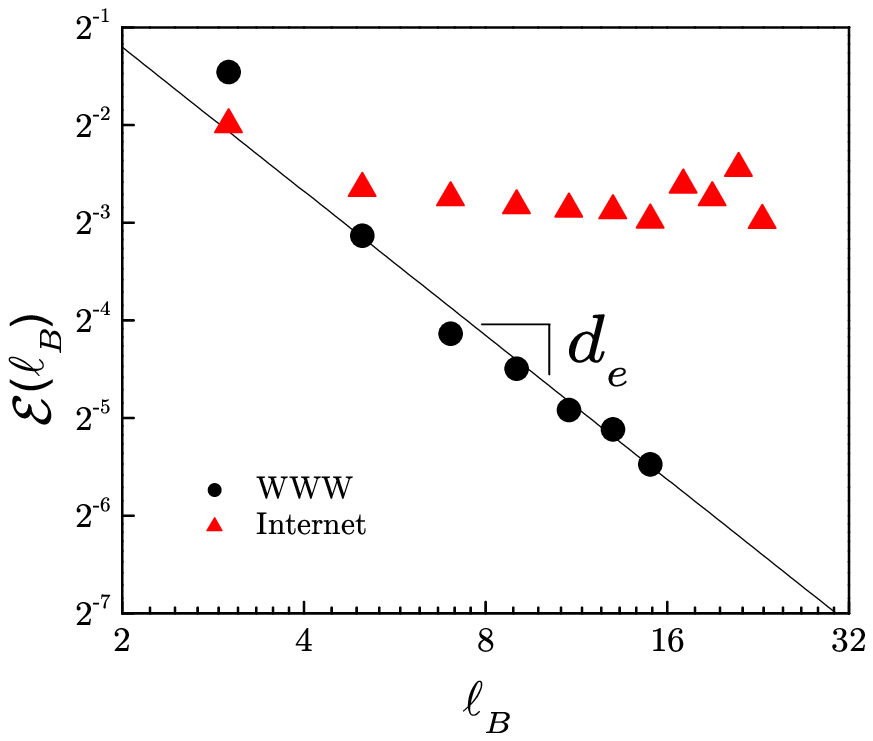}}} } \caption {} \label{kk}
\end{figure*}

{\bf Growth mechanism.---} The renormalization scheme tiles a
network of $N$ nodes with $N_B(\ell_B)$ boxes using the
box-covering algorithm \cite{shm}, as shown in Fig.
\ref{renormalization}a.
 The boxes contain nodes separated by a distance
$\ell_B$, measured as the length of the shortest path between
nodes. Each box is subsequently replaced by a node, and the
process is repeated until the whole network is reduced to a single
node. The way to distinguish between fractal and non-fractal
networks is represented in their scaling properties as seen in
Fig. \ref{kk}a and \ref{kk}b. Fractal networks can be
characterized by the following scaling relations (Fig. \ref{kk}a):
\begin{equation}
\begin{split}
N_B(\ell_B)/N\sim\ell_B^{-d_B} ~~~~ \textrm{and} ~~~~
k_B(\ell_B)/k_{hub}\sim \ell_B^{-d_k},
\end{split}
\label{fractal}
\end{equation}
where $k_{hub}$ and $k_B(\ell_B)$ are the degree of the most
connected node inside each box and that of each box respectively
(Fig. \ref{renormalization}a). Although both of them are partial
variables, the ratio between them is a global quantity, depending
only on the length scale $\ell_B$, as we showed in \cite{shm}. The
two exponents $d_B$ and $d_k$ are the fractal dimension and the
degree exponent of the boxes, respectively. While the term
"fractal dimension" is usually reserved for geometrical
self-similarity, here we relax the usage to include the
topological self-similarity as well. For a non-fractal network
like the Internet (Fig. \ref{kk}b), we have $d_B\to\infty$ and
$d_k\to\infty$; the scaling laws in Eq. (\ref{fractal}) are
replaced by exponential functions.

Based on the results leading to Eq. (\ref{fractal}), we propose a
network growth dynamics as the inverse of the renormalization
procedure. Thus, the coarse-grained networks of smaller size are
network structures appearing earlier in time, as exemplified in
Fig. \ref{renormalization}a. A present time network with
$\tilde{N}(t)$ nodes is tiled with $N_B(\ell_B)$ boxes of size
$\ell_B$. Each box represents a node in a previous time step, so
that $\tilde{N}(t-1)=N_B(\ell_B)$. The maximum degree of the nodes
inside a box corresponds to the present time degree: $\tilde{k}(t)
= k_{hub}$, which is renormalized such that
$\tilde{k}(t-1)=k_B(\ell_B)$. The tilde over the quantities are
needed in order to differentiate the dynamical quantities, such as
the number of nodes as a function of time, $\tilde{N}(t)$, from
the static quantities, such the number of nodes of the present
network, $N$, or the number of nodes of the renormalized network,
$N_B$. The renormalization procedure applies to many complex
networks in Nature \cite{shm}. These includes fractal networks
such as WWW, protein interaction networks of E. {\it coli}, the
yeast \cite{yeast} and human, and metabolic networks of 43
different organisms from the three domains of life, and some
sociological networks. The renormalization scheme can be applied
to non-fractal networks, such as the Internet, as well. Below we
will show that the main difference between these two groups is in
the connectivity correlation. We also provide empirical,
analytical and modelling evidences supporting this theoretical
framework based on the validity of exponents, scaling theory, and
statistical properties of the connectivity correlation.

{\bf Correlation.---} A question of importance to elucidate the
selection rules governing the fractality of the network is to
determine how the nodes in older networks are connected to those
of the present day. The answer lies in the statistical property of
correlation between the nodes and boxes within a network
configuration. Studying the correlation profile in real networks
similar to those  considered in \cite{maslov,vespignani2,newman2}
provides initial hints to the above question. The correlation
profile \cite{maslov} compares the joint probability distribution,
$P(k_1,k_2)$, of finding a node with $k_1$ links connected to a
node with $k_2$ links with their random uncorrelated counterpart,
$P_r(k_1,k_2)$, which is obtained by random swapping of the links,
yet preserving the degree distribution. A plot of the ratio
$R(k_1,k_2) = P(k_1,k_2)/P_r(k_1,k_2)$ provides evidence of
correlated topological structure that deviates from the random
uncorrelated case.

At first glance, a qualitative classification based on the
strength of the anticorrelation of different networks can be
obtained by normalizing the ratio $R(k_1,k_2)$ to that of a given
network, for instance the WWW \cite{barabasi1999}, (Supplementary
Materials, Section \ref{uncorrelated}). Figure \ref{kk}c and
\ref{kk}d show the correlation profiles of the cellular metabolic
network of E. {\it coli} \cite{cellular}, which is known to be
fractal, and the Internet at the router level \cite{lucent}, which
has a non-fractal topology. The fractal network poses a higher
degree of anticorrelation or disassortativity;
 nodes with a large degree tend to be connected
with nodes of a small degree. On the other hand, the non-fractal
Internet is less anticorrelated. Thus, fractal topologies seem to
display a higher degree of hub repulsion in their structure than
non-fractals. However, for this property to be the  hallmark of
fractality, it is required that the anticorrelation appears not
only in the original network (captured by the correlation profiles
of Fig. \ref{kk}c and Fig. \ref{kk}d), but also in the
renormalized networks at all length scales. We note that other
measures of anticorrelation, such as the Pearson coefficient $r$
of the degrees at the end of an edge \cite{newman2}, cannot
capture the difference between fractal and non-fractal network. We
find that $r$ is not invariant under renormalization.

{\bf Mathematical model.---} To quantitatively link the
anticorrelation at all length scales to the emergence of
fractality, we next develop a mathematical framework and
demonstrate the mechanism for fractal network growth. In the case
of modular networks, stemming from Eqs. (\ref{fractal}), we
require that
\begin{equation}
\begin{split}
\tilde{N}(t) = n \tilde{N}(t-1),\\
\tilde{k}(t)=s \tilde{k}(t-1),\\
\tilde L(t) + L_0= a (\tilde L(t-1)+L_0),\\
\end{split}
\label{exp}
\end{equation}
where $n>1$, $s>1$ and $a>1$ are time-independent constants and
$\tilde L(t)$ is the diameter of the network defined by the
largest distance between nodes. The first equation is analogous to
the multiplicative process naturally found in many population
growth systems \cite{vankampen}. The second relation is analogous
to the preferential attachment rule \cite{ba}. It gives rise to
the scale-free probability distribution of finding a node with
degree $k$, $P(k)\sim k^{-\gamma}$. The third equation describes
the growth of the diameter of the network and determines whether
the network is small-world \cite{watts} and/or fractal. Here we
introduce the characteristic size $L_0$, the importance of which
lies in describing the non-fractal networks. Since every quantity
increases by a factor of $n$, $s$ and $a$, we first derive
(Supplementary Materials Section \ref{theory}) the scaling
exponents in terms of the microscopic parameters: $d_B = \ln n/\ln
a,$ $d_k = \ln s/\ln a.$ The exponent of the degree distribution
satisfies $\gamma = 1+\ln n/\ln s$. The dynamics represented by
Eqs. (\ref{exp}) consequently leads to a modular structure where
modules are represented by the boxes. While modularity has often
been identified with the scaling of the clustering coefficient
\cite{ravasz}, here we propose an alternative definition of
``modular network" as the one whose statistical properties remain
invariant (in particular, an invariant degree distribution with
the same exponent $\gamma$, see Supplementary Materials Section
\ref{invariant}) under renormalization.

In order to incorporate different growth modes in the dynamical
Eqs. (\ref{exp}) we consider, without loss of generality, two
modes of connectivity between boxes, whose relative frequencies of
occurrence are controlled by the probability $e$ representing the
hub-hub attraction. {\it (i)} Mode I with probability $e$ (Fig.
\ref{renormalization}b): two boxes are connected through a direct
link between their hubs leading to hub-hub attraction. {\it (ii)}
Mode II with probability $1-e$ (Fig. \ref{renormalization}c):
 two boxes are connected via non-hubs
leading to hub-hub repulsion  or anticorrelation. We will show
that Mode I leads to non-fractal networks while Mode II leads to
fractal networks. In practice, though Eqs. (\ref{exp}) are
deterministic, we combine these two modes according to the
probability $e$, which renders our model probabilistic.

Formally, for a node with $\tilde{k}(t-1)$ links at time $t-1$, we
define $\tilde{n}_{h}(t)$ as the number of links which are
connected to hubs in the next time step (see Fig.
\ref{renormalization}a). Then the probability $e$ satisfies:
\begin{equation}
\tilde{n}_{h}(t) = e ~ \tilde{k}(t-1). \label{e}
\end{equation}

Using the analogy between time evolution and renormalization, we
introduce the corresponding quantity, $n_h(\ell_B)$, and defines
the ratio ${\cal E}(\ell_B)\equiv n_{h}(\ell_B) / k_B(\ell_B)$.
The nonlinear relation between $t$ and $\ell_B$ leads to the
$\ell_B$ dependence on $\cal E$ (see Supplementary Materials,
Section \ref{theory}). In the extreme case of strong hub
attraction, where the hubs of the boxes are connected at all
length scales, we have ${\cal E}(l_B)\sim $ constant. On the other
hand, hub repulsion leads to decreasing ${\cal E}(\ell_B)$ with
$\ell_B$. From scaling we obtain a new exponent $d_e= -\ln e/\ln
a$ characterizing the strength of the anticorrelation in a
scale-invariant way:
\begin{equation}
{\cal E}(\ell_B)\sim \ell_B^{-d_e}. \label{E}
\end{equation}
Fig. \ref{kk}e shows ${\cal E}(\ell_B)$ for two real fractal and
non-fractal networks: a map of the WWW domain (nd.edu) consisting
of 352,728 web-sites \cite{barabasi1999} and  a map of the
Internet at the router level consisting of  284,771 nodes
\cite{lucent}. We find that for the fractal WWW,  $d_e=1.5$,
indicating that it exhibits strong anticorrelation. On the other
hand, the non-fractal Internet shows ${\cal E}(\ell_B)\sim$
constant.

These results confirm that fractal networks, including the protein
interaction network \cite{dip} (with $d_e=1.1$) and the metabolic
network of E. {\it coli} \cite{cellular} (with $d_e=4.5$), do have
strong hub repulsion at all length scales and non-fractal networks
have no or weak hub repulsion.

A general limitation when analyzing the scaling behavior of
complex networks is the small range in which the scaling is valid.
This is due to the  small-world property that restricts the range
of $\ell_B$ in Fig. \ref{kk}. As an attempt to circumvent this
limitation, we offer not only the empirical determination of the
exponents but also scaling theory and models where the exponents
can be further tested. We should also point out that large
exponents (such as $d_e=4.5$ for E. {\it coli}) may not be
distinguishable from exponential behavior (infinite exponent). In
this case, however the large exponent $d_e$ for E. {\it coli}
agrees with our theoretical framework, since it corresponds to a
network with large anticorrelation in the connectivity and the
subsequent small fractal dimension. In terms of the model, this
corresponds to the limit of $e\rightarrow 0$.

Next we show how the different growth modes reproduce the
empirical findings. While each mode leads to the scale-free
topology, they differ in their fractal and small-world properties.
Mode I alone $(e=1)$ exhibits the small-world effect, but is not
fractal due to its strong hub-hub attraction (see Fig.
\ref{renormalization}b). On the other hand, Mode II alone ($e=0$,
Fig. \ref{renormalization}c) gives rise to a fractal network.
However, in this case, the anticorrelation is strong enough to
push the hubs far apart, leading to the disintegration of the
small-world. Full details of the implementation of Mode I and Mode
II are given in the Supplementary Materials Section \ref{minimal}
and \ref{small-world}.

\begin{figure*}
\centerline{ \hbox{ (a)\resizebox{7cm}{!} {
\includegraphics{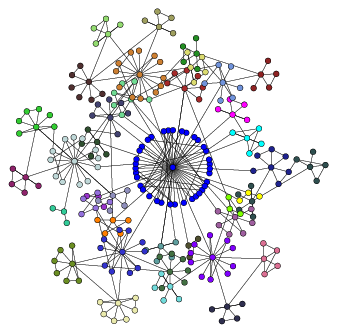}} (b)\resizebox{7cm}{!} {
\includegraphics{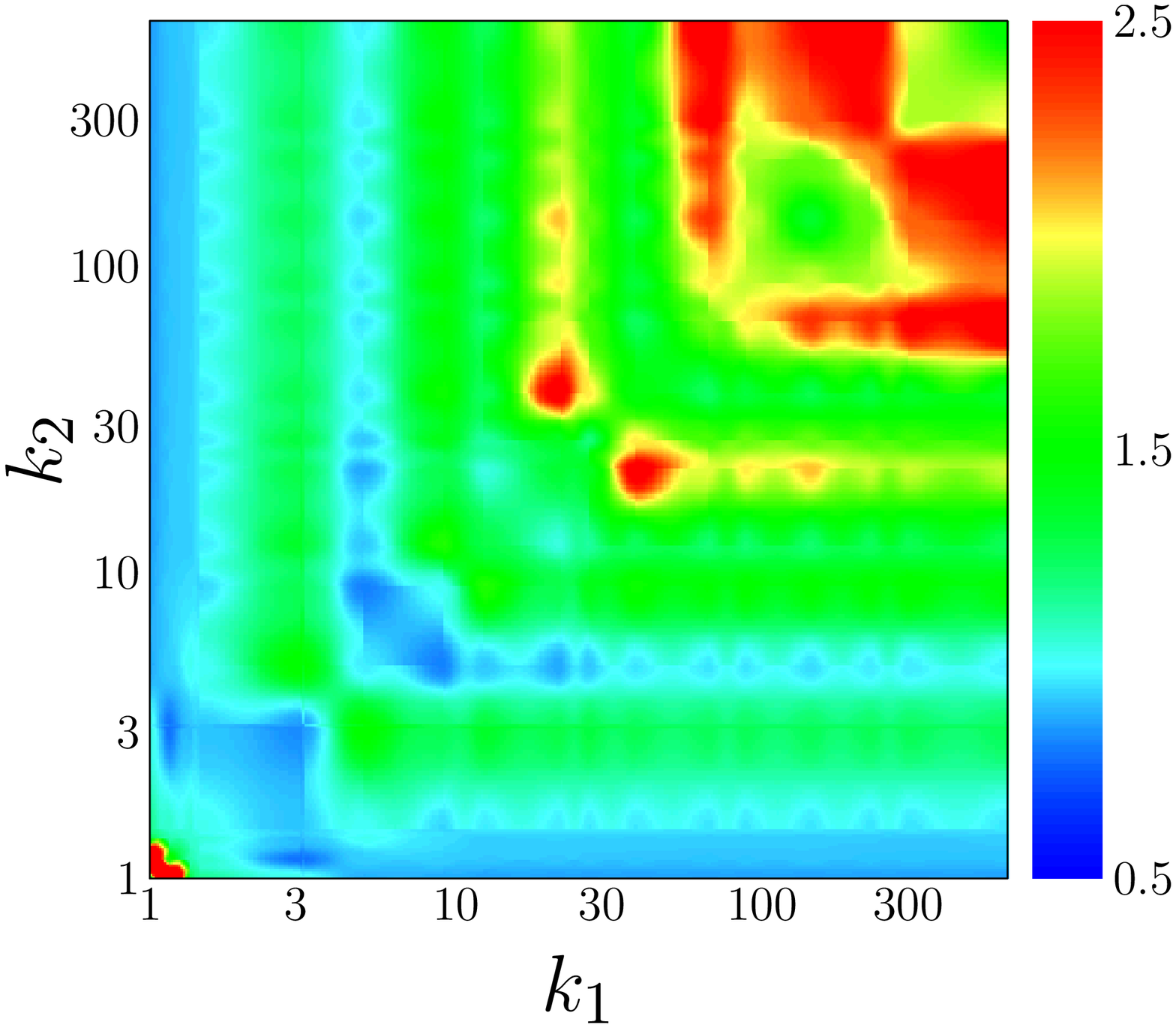}} }} \centerline{ \hbox{
(c)\resizebox{8cm}{!} { \includegraphics{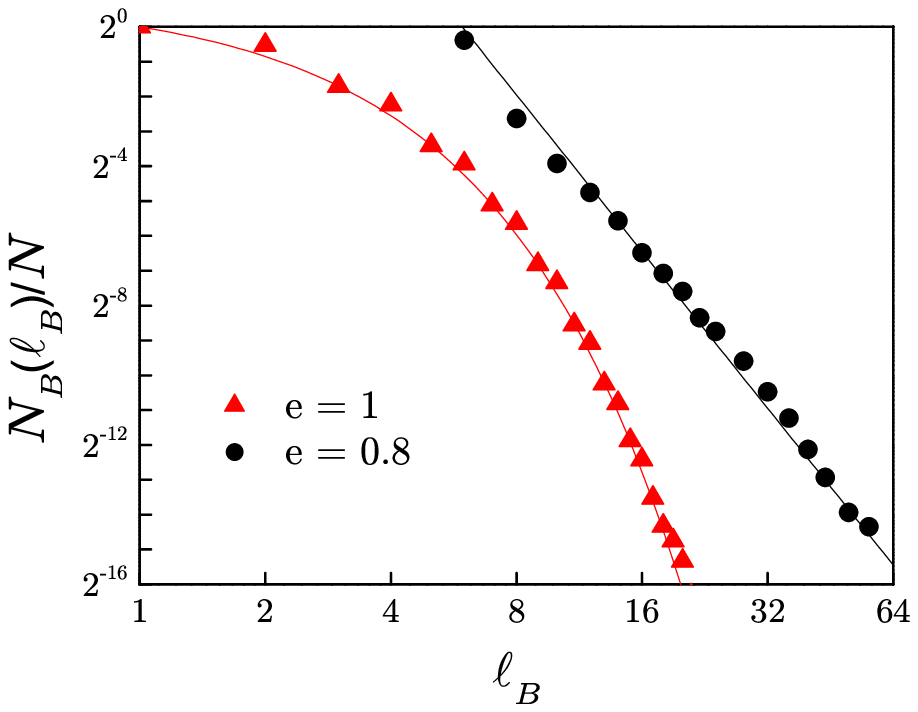}}
(d)\resizebox{8cm}{!} { \includegraphics{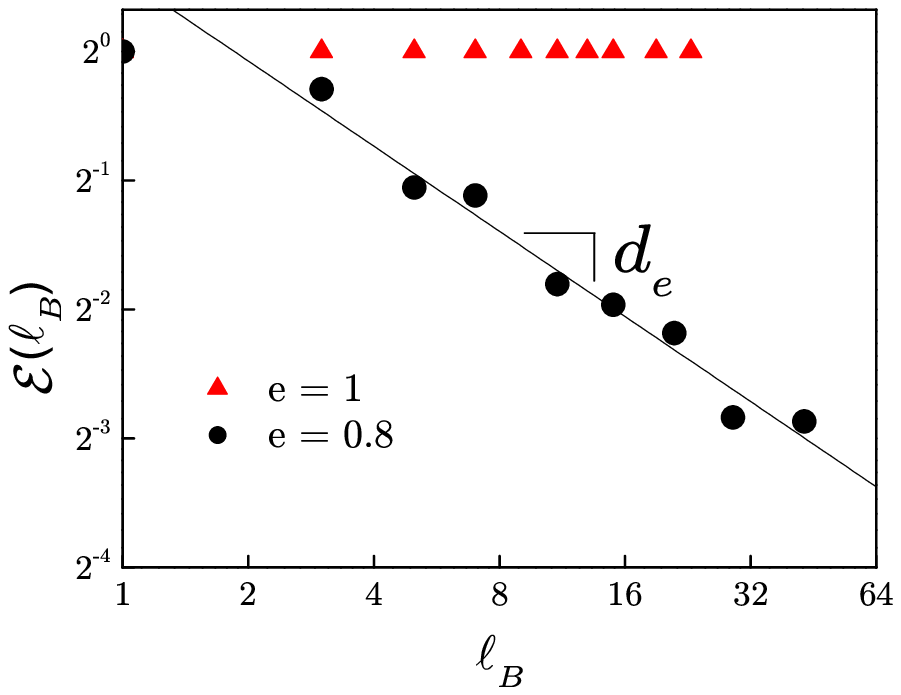}} }}
\caption { } \label{model}
\end{figure*}

These results suggest that the simultaneous appearance of both the
small-world and fractal properties in scale-free networks is due
to a combination of the growth modes. In general, the growth
process is a stochastic combination of Mode I (with probability
$e$) and Mode II (with probability $1-e$). For the intermediate
($0<e<1$), the model predicts finite fractal exponents  $d_B$ and
$d_k$ and also bears the small-world property due to the presence
of Mode I. Such a fractal small-world and scale-free network is
visualized in Fig. \ref{model}a for $e=0.8$. Supporting evidences
are given by  {\it (i)} Fig. \ref{model}b, which shows that the
model with $e=0.8$ is more anticorrelated than the $e=1$ model
(Mode I), {\it (ii)} Fig. \ref{model}c, which shows the power law
dependence of $N_B$ on $\ell_B$ for the fractal structure
($e=0.8$), and the exponential dependence of the non-fractal
structure ($e=1$), and {\it (iii)} Fig. \ref{model}d showing that
Mode I reproduces ${\cal E}(\ell_B) \sim$ constant while the
$e=0.8$ model gives  ${\cal E}(\ell_B) \sim \ell_B^{-d_e}$, which
is in agreement with the empirical findings of Fig. \ref{kk}e on
real networks (the exponent $d_e=-\ln 0.8/\ln 1.4 = 0.66$ is
predicted by the analytical formula according to Supplementary
Material \ref{theory}). Furthermore, in the Supplementary
Materials Section \ref{minimal} we show that the predicted
scale-free distribution is invariant under renormalization.
Although simplistic, this minimal model clearly captures an
essential property of networks: the relationship between
anticorrelation and fractality (see Methods for more details). We
have also considered the contribution of loops, which we find does
not change the general conclusions of this study.

\begin{figure*}
\hbox{ \resizebox{17cm}{!} { \includegraphics{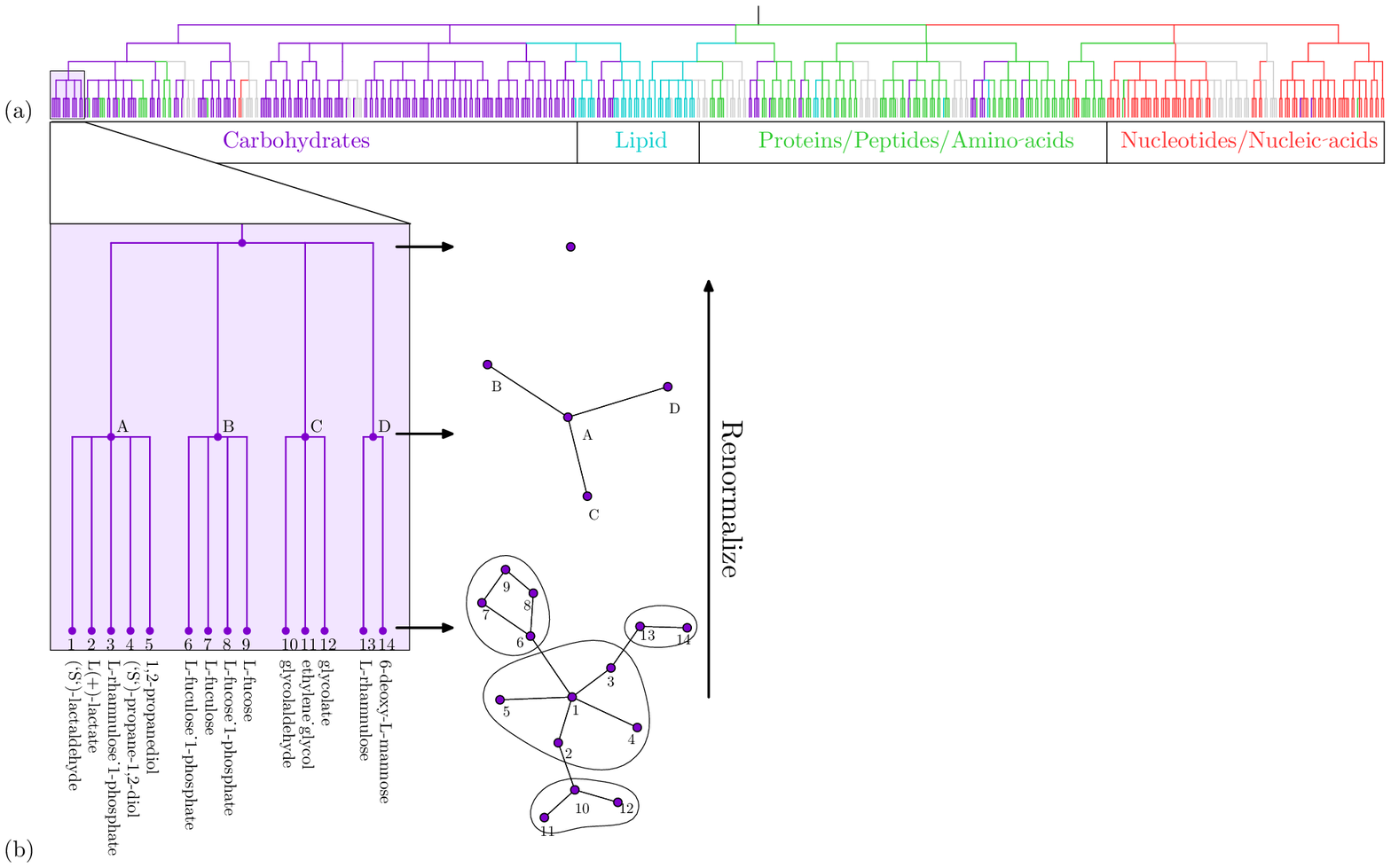}}
\hspace{-8cm} (c) \resizebox{8cm}{!} {
\includegraphics{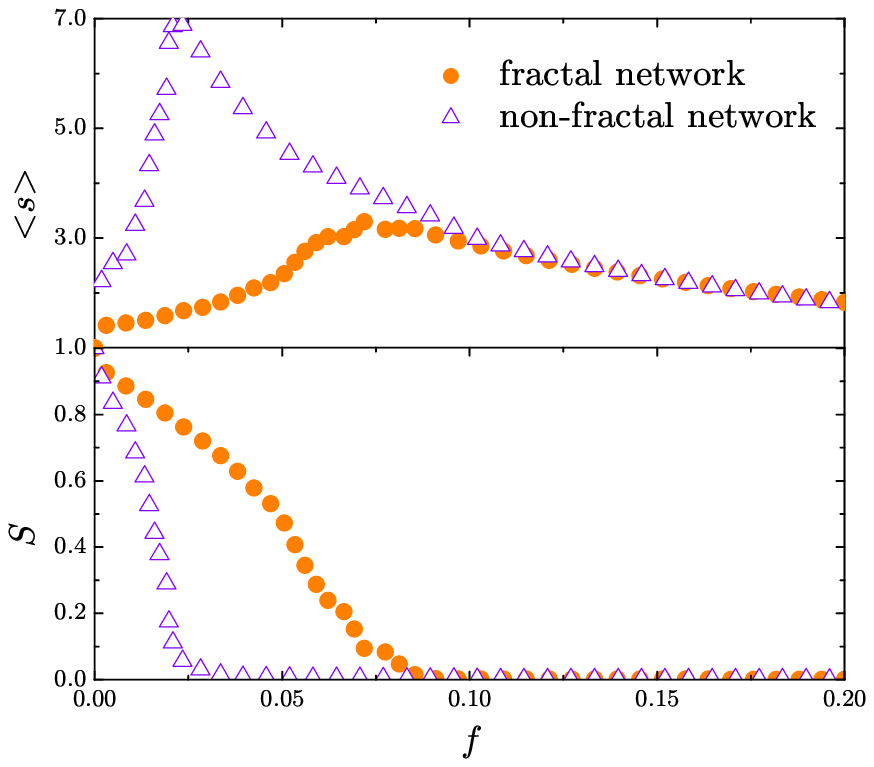}} } \caption {} \label{modular}
\end{figure*}

{\bf Modularity.---} The scale-invariant properties naturally lead
to the appearance of a hierarchy of self-similar nested
communities or modules. In this novel point of view, boxes
represent nested modules of different length scales. The
importance of modular structures is stressed in biological
networks, where questions of function and evolutionary importance
are put to the test \cite{hopfield,ravasz,newman3,nested}. The
relevant question is whether the self-similar hierarchy of boxes
encodes the information about the functional modules in biological
networks. To answer this question we analyze the fractal metabolic
network of E. {\it coli} \cite{cellular} which has been previously
studied using standard clustering algorithms \cite{ravasz}. Here
we show that by repeatedly applying the renormalization we produce
a tree with branches that are closely related to the biochemical
annotation, such as carbohydrates, lipids, amino acid, etc
\cite{ravasz}. We renormalize the network at a given box size and
cluster the substrates which belong to the same box and repeat the
procedure to generate the hierarchical tree shown in Fig.
\ref{modular}a. In Fig. \ref{modular}b (the right-bottom scheme),
we see a subnet of the original metabolic network with 14 nodes.
They correspond to the bottommost layer of the hierarchical tree
in the left. The box covering with $\ell_B=3$ indicates that this
subnet contains four modules. The coarse-grained network is shown
in the right-middle with 4 nodes: A, B, C and D. The next stage of
renormalization combines these four nodes to one single node or
class. Following this algorithm, we coarse-grain the network and
classify the nodes at different levels. In Fig. 5a, we show this
classification for the entire metabolic network. The different
colors correspond to distinct functional modules, as we annotate
in the bottom of the tree (carbohydrates, lipids, etc.). The clear
division of biological functions in the hierarchical tree suggests
that the metabolic network is organized in a self-similar way.

The main known biochemical classes of the substrates emerge
naturally from the renormalization tree, indicating that the boxes
capture the modular structure of the metabolic  network of E. {\it
coli}. The same analysis  reproduces the modular structure of the
protein interaction network of the yeast further suggesting the
validity of our analysis \cite{yeast}.

{\bf Robustness.---} Finally our results suggest the importance of
self-similarity in the evolution of the topology of networks.
Understanding the growth mechanism is of fundamental importance as
it raises the question of its motivation in Nature.  For instance,
given that systems in biology are fractal, there could  be an
evolutionary drive for the creation of such networks.  A parameter
relevant to evolution is the robustness of the network, which can
be compared between fractal and non-fractal networks.

Non-fractal scale-free networks, such as the Internet, are
extremely vulnerable to targeted attacks on the hubs
\cite{barabasi-attack}. In such non-fractal topologies, the hubs
are connected and form a central compact core (as seen in Fig.
\ref{kk}b), such that the removal of few largest hubs (those with
the largest degree) has catastrophic consequences for the network
\cite{barabasi-attack,kitano}. Here we show that the fractal
property of networks significantly increases the robustness
against targeted attacks since the hubs are more dispersed in the
network (see Fig. \ref{kk}a). Figure \ref{modular}c shows a
comparison of robustness between a fractal and non-fractal
network. The comparison is done between model networks of the same
$\gamma=2.8$, the same number of nodes (74,000), the same number
of links, the same amount of loops and the same clustering
coefficient (see Supplementary Materials Section \ref{attack}).
Thus the difference in the robustness seen in this figure is
attributed solely to the different degree of anticorrelation. We
plot the relative size of the largest cluster, $S$, and the
average size of the remaining isolated clusters, $\langle s
\rangle$, after removing  a fraction $f$ of the largest hubs for
both networks \cite{barabasi-attack}. While both networks collapse
at a finite fraction $f_c$, evidenced by the decrease of $S$
toward zero and the peak in $\langle s \rangle$, the fractal
network has  a significantly larger threshold ($f_c\approx 0.09$)
compared to the non-fractal threshold ($f_c\approx 0.02$)
suggesting a significantly higher robustness of the fractal
modular networks to failure of the highly connected nodes. This
could explain why evolutionary constraints on  biological networks
have led to fractal architectures. It is important to note that
the comparison in Fig. \ref{modular}c is between two networks
which preserve the modularity. Our results should be understood as
follows: given that a network has a modular structure, then the
most robust network is the one with fractal topology. There are
other ways to increase robustness by, for instance, fully
connecting the hubs in a central core \cite{doyle}, but this
arrangement does not preserve the modularity.

{\bf Summary.---} We find that the statistical properties of many
real networks are well consistent with the predictions of the
proposed multiplicative model. Networks that can be captured by
our theoretical framework include fractal networks (WWW, protein
interactions, metabolic networks and some collaboration networks)
and the non-fractal networks such as the Internet. The validity of
the proposed framework is supported by the predicted scaling
exponents ($d_B$, $d_k$, $d_e$) in many real networks as well as
the general properties of the connectivity distribution captured
by $P(k_1,k_2)$ and the scaling relationships predicted by the
multiplicative growth process of our model. Our results
demonstrate that nodes are organized around dispersed hubs in
self-similar nested modules \cite{nested} characterized by
different functionalities. These then compartmentalize the hubs
\cite{maslov}, and protect them from a failure at the system level
\cite{kitano}. Hence, these modules function relatively
autonomously so that a failure in one module cannot propagate
easily to the next.  This may provide a significantly higher
protection against intentional attacks reducing the high
vulnerability--- the Achilles' heel--- of non-fractal scale-free
networks.

\clearpage

\centerline{\bf METHODS}

{\bf Details of the Mode I and Mode II of growth in the minimal
model.---} Mode I: To each node with degree $\tilde{k}(t-1)$ at
time $t-1$, $m \tilde{k}(t-1)$ offspring nodes are attached at the
next time step ($m=2$ in the example of Fig.
\ref{renormalization}b. As a result we obtain a scale-free
non-fractal network: $N_B(l_B)/N \sim$$ \exp(-\frac{\ln
n}{2}\ell_B)$ and $k_B(l_B)/k_{hub}\sim \exp(-\frac{\ln
s}{2}\ell_B)$, implying that both exponents $d_B$ and $d_k$ are
infinite (since $a\rightarrow 1$ then $d_B=\ln n/\ln a \rightarrow
\infty$ and $d_k=\ln s/\ln a \rightarrow \infty$). This is a
direct consequence of the linear growth of the diameter $\tilde
L(t)$. Moreover, the additive growth in the diameter with time
implies that the network is small-world. This mode is similar to a
class of models called pseudo-fractals \cite{doro,jkk}. Mode II:
It gives rise to a fractal topology but with a breakdown of the
small-world property. The diameter increases multiplicatively
leading to an exponential growth with time, and consequently to a
fractal topology with finite $d_B$ and $d_k$.

\clearpage

FIG. \ref{renormalization}. Self-similar dynamical evolution of
networks. (a) The dynamical growth process can be seen as the
inverse renormalization procedure with all the properties of the
network being invariant under time evolution. In this example
$\tilde{N}(t)=16$ nodes are renormalized with $N_B(\ell_B)=4$
boxes of size $\ell_B=3$. (b) Analysis of Mode I, only: the boxes
are connected directly leading to strong hub-hub attraction or
assortativity. This mode produces a scale-free, small-world
network but without the fractal topology. (c) Mode II alone
produces a scale-free with a fractal topology but not the
small-world effect. Here the boxes are connected via non-hubs
leading to hub-hub repulsion or disassortativity.

FIG. \ref{kk}. Empirical results on real complex networks. (a)
Schematics showing that fractal networks are characterized by a
power law dependence between $N_B$ and $\ell_B$ while (b)
non-fractal networks are characterized by an exponential
dependence. (c) Plot of the correlation profile of the fractal
metabolic  network of E. {\it coli}, $R_{{\rm E.}{\it
coli}}(k_1,k_2)/R_{\rm WWW}(k_1,k_2)$, and (d) the non-fractal
Internet $R_{\rm Int}(k_1,k_2)/R_{\rm WWW}(k_1,k_2)$, compared
with the profile of the WWW in search of a signature of
fractality. (e) Scaling of $\cal{E}(\ell_B)$ as defined in Eq.
(\ref{E}) for the fractal topology of the WWW with $d_e=1.5$, and
the non-fractal topology of the Internet showing that fractal
topologies are strongly anticorrelated at all length scales. In
order to calculate $\cal E$ (and in all the calculations in this
study) we tile the network by first identifying the nodes which
are the center of the boxes with the largest mass and sequentially
centering the boxes around these nodes.

FIG. \ref{model}. Predictions of the renormalization growth
mechanism of complex networks. (a) Resulting topology predicted by
the minimal model  for $e=0.8$, $n=5$, $a=1.4$, $s=3$ and $m=2$.
The colors of the nodes show the modular structure with each color
representing a different box. We also include loops in the
structure as discussed in the Supplementary Materials, Section
\ref{attack}. (b) Ratio $R_{e=1}(k_1,k_2)/R_{e=0.8}(k_1,k_2)$ to
compare the hub-hub correlation emerging from the model networks
generated with $e=1$ and $e=0.8$, respectively. (c) Plot of $N_B$
versus $\ell_B$ showing that Mode I is non-fractal (exponential
decay) and $e=0.8$ is fractal (power-law decay) according to (b)
and in agreement with the empirical results of Fig. \ref{kk}. (d)
Scaling of ${\cal E}(\ell_B)$  reproducing the behavior of fractal
networks for $e=0.8$ and non-fractal networks Mode I, $e=1$,
 as found empirically in Fig. \ref{kk}e.

FIG. \ref{modular}. Practical implications of the renormalization
growth approach and fractality. (a) Renormalization tree of the
metabolic network of E. {\it coli} leading to the appearance of
the functional modules. The colors of the nodes and branches in
the tree denote the main biochemical classes as: carbohydrates,
lipids, proteins, peptides and aminoacids, nucleotides and nucleic
acids, and coenzymes and prosthetic groups biosynthesis (grey).
(b) Details of the construction of three levels of the
renormalization tree for $\ell_B=3$ for 14 metabolites in the
carbohydrate biosynthesis class as shown in the shaded area in
(a). (c) Vulnerability under intentional attack of a
 non-fractal network generated by Mode I ($e=1$) and a fractal network
generated by Mode II ($e=0)$. The  plot shows the relative size of
the largest cluster, $S$, and the average size of the remaining
isolated clusters, $\langle s \rangle$ as a function of the
removal fraction $f$ of the largest hubs for both networks.

{\small {\bf Supplementary Information} accompanies the paper on
www.nature.com/naturephysics}

{\small {\bf Correspondence} and requests for materials should be
addressed to H. A. M (makse$@$mailaps.org).}

{\small {\bf Acknowledgements}. We would like to thank J. Bruji\'c
for illuminating discussions and E. Ravasz for providing the data
on the metabolic network. SH wishes to thank the Israel Science
Foundation, ONR and Dysonet for support. This work is supported by
the National Science Foundation, DMR-0239504 to HAM. }

\clearpage

\centerline{\bf SUPPLEMENTARY MATERIAL}

\vspace{1cm}

The Supplementary Material section is organized as follows:
Section \ref{modeling} shows that the available scale-free models
are non-fractals. Section \ref{uncorrelated} describes more
results on the correlation supplementing the various analysis and
calculations in the main text.Section \ref{invariant} reveals that
the statistical properties, in particular, the degree distribution
keeps invariant under renormalization for real-world networks.
Section \ref{theory} explains in detail the various theoretical
results presented in the main text and elaborates on the
extensions of the minimal model of fractal growth. Finally,
Section \ref{attack} gives more details of the study of the
robustness of fractal and non-fractal networks under intentional
attack.

\section{Study of scale-free  models}
\label{modeling}

While the origin of the scale-free property can be reduced to two
basic mechanisms: growth and preferential attachment, as
exemplified by the Barab\'asi-Albert model (BA model \cite{ba}),
the  empirical result of fractality cannot be explained only in
those terms. Notice that the term "scale-free" coined by
Barab\'asi-Albert \cite{ba} refers to the absence of a typical
number of links, as exemplified by a power-law distribution of
degree connectivity, but it does not refer to the length scale
invariance found in \cite{shm}.

\begin{figure}
\centerline{ (a) \resizebox{7cm}{!} {
 \includegraphics{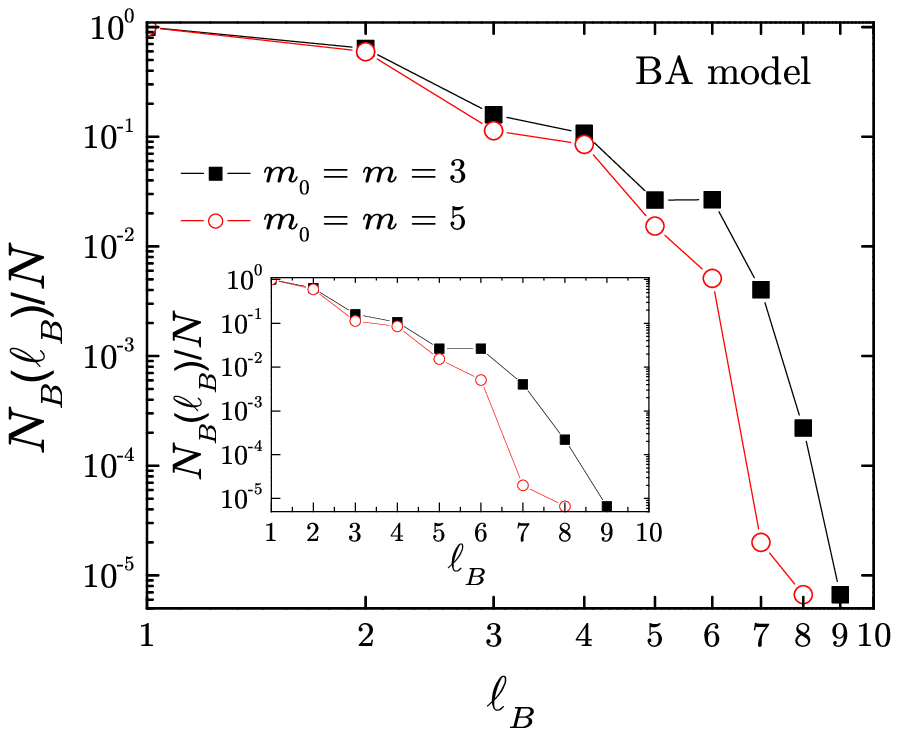}}
(b) \resizebox{7cm}{!} {
 \includegraphics{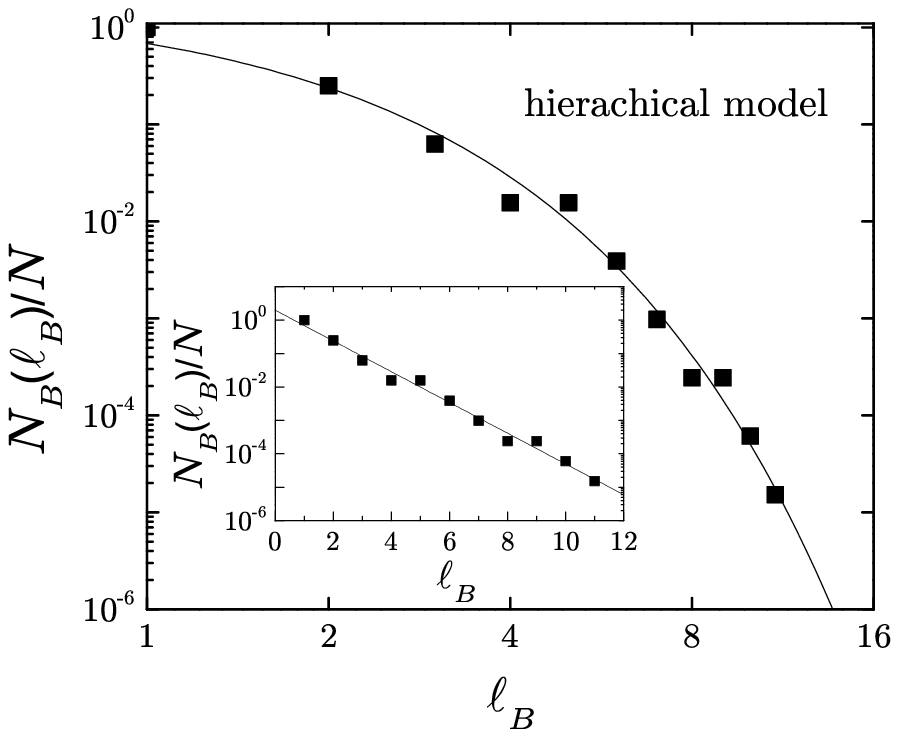}}
} \centerline{ (c) \resizebox{7cm}{!} {
 \includegraphics{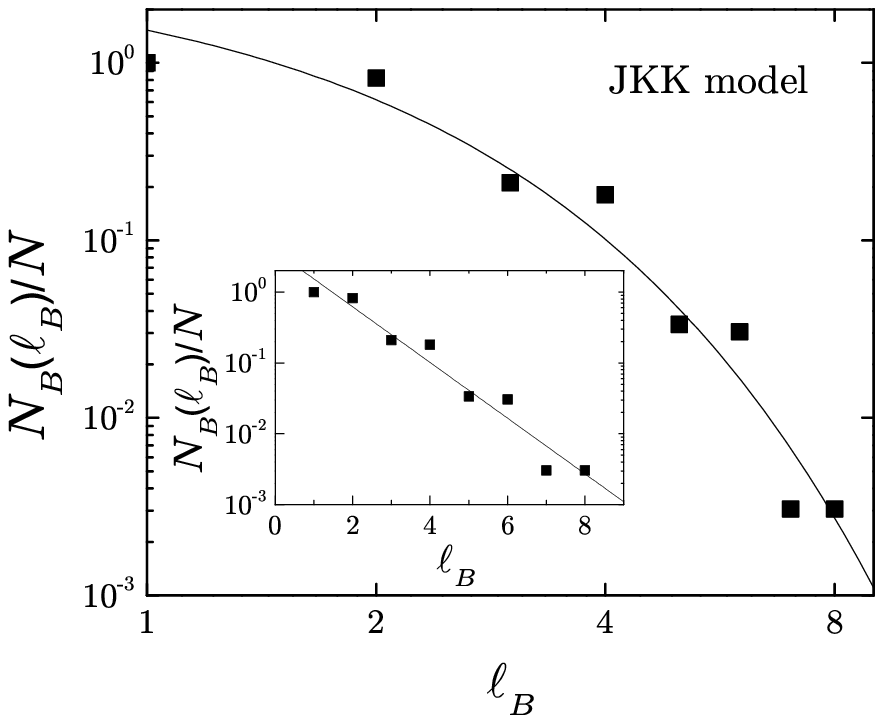}}
} \caption {Test of fractality in scale-free models. Number of
boxes versus size of the boxes of different models showing that
they do not follow a power-law, but and exponential (or faster)
function as shown in the insets. Thus, they are non-fractal. (a)
The BA model of preferential attachment \cite{ba}, (b) the
hierarchical model \cite{ravasz} and (c) the model of Jung, Kim,
and Kahng (JKK model)\cite{jkk} which is an example of pseudo
fractal models as discussed by Dorogovtsev and Mendes
\cite{doro}.} \label{models}
\end{figure}

We find that all  models of scale-free networks such as the BA
model of preferential attachment \cite{ba}, the hierarchical model
\cite{ravasz}, and the so-called pseudo fractal models and trees
\cite{doro,jkk} are non-fractals. In Fig. \ref{models} we plot the
number of boxes $N_B$ versus $\ell_B$ for the models showing that
in all the cases the decay of $N_B(\ell_B)$ is exponential or
faster, indicating either an infinite $d_B$ or not a well-defined
fractal dimension.

In the present study we find the relation $\gamma = 1+ \ln n/\ln
s$, by using $d_B = \ln n/\ln a,$ and $d_k = \ln s/\ln a$, as
explained in the text. However, non-fractal networks satisfy this
relation as well despite the infinite fractal dimension
$d_B\rightarrow \infty$. Thus in general we say that when $\gamma
= 1+ \ln n/\ln s$ is satisfied, then the degree distribution is
invariant under renormalization.

\section{The random uncorrelated scale-free model and the correlation
profiles} \label{uncorrelated}

It is instructive to analyze the degree of correlation in networks
considering the deviations of the joint probability distribution
$P(k_1,k_2)$ from the random uncorrelated scale-free case
$P_r(k_1,k_2)$. This latter model is obtained by, for instance,
random swapping of the links in a given network \cite{maslov}, so
that the degree distribution is preserved, but the correlation is
completely lost.

The study of the ratio $R(k_1,k_2) = P(k_1,k_2)/P_r(k_1,k_2)$
reveals that most of the networks such as metabolic and protein
interaction networks, the Internet and WWW are anticorrelated in
comparison with the uncorrelated random case. This is because,
even though this model is uncorrelated, there is still an
effective attraction between the hubs since there is a large
probability to randomly connect two nodes with large degrees.
Thus,  a plot of the ratio $R = P(k_1,k_2)/P_r(k_1,k_2)$ reveals
that most of the real networks are anticorrelated in comparison
with the uncorrelated model. Therefore this ratio does not allow
to distinguish between fractal and non-fractal networks.

In search of uncovering the extent of anticorrelation that are
needed to obtain fractals we study the ratios for different
networks by using the WWW as a reference (the use of any other
network as a reference would lead to the same conclusions). This
is done in the main text in Figs. \ref{kk}c and \ref{kk}d and for
the model in Fig. \ref{model}b. These plots  should be interpreted
as follows: For instance, in Fig. \ref{kk}c, let us take a large
degree $k_1=100$ as an example. Then we see that the ratio
$R_{E.coli}(k_1,k_2)/R_{WWW}(k_1,k_2)$ has a maximum for
$k_2\approx 5$ (red-yellow scale) for small $k_2$ ($k_2 < 10$),
and a minimum (blue scale) for large $k_2
> 10$. This means that the metabolic network
has less probability to have hub-hub connection (two nodes with
large degree connected) than a hub-non hub connection, when
compared to the WWW. Therefore the metabolic network of E.coli is
more anticorrelated than the WWW. In the same way, Fig. \ref{kk}d
shows that the hubs in the Internet have more probability to
connect with other hubs than in the WWW, and therefore the
Internet is less anticorrelated than the WWW. Therefore these
patterns reveal that the fractal cellular networks are strongly
anticorrelated (dissortative).

The same analysis is performed for the model in Fig. \ref{model}b
in the main text. For instance, in this figure, given a large
degree $k_1=300$, the ratio $R_{e=1}(k_1,k_2)/R_{e=0.8}(k_1,k_2) $
is small (blue/green region) for small $k_2$ ($k_2 < 10$) but
large (red/yellow region) for large $k_2$ ($k_2 > 10$). This means
that the network with $e=1$ is more likely to have hub-hub
connections than the $e=0.8$ case. Thus, the profile shows how
$e=0.8$ is more anticorrelated than Mode I.

In summary, using a short notation, the strength of hubs repulsion
satisfies: Uncorrelated scale-free model $<$ Internet $<$ WWW $<$
protein interaction $<$ metabolic networks. For the model we have:
Mode I $<$ Model with intermediate $e <$ Mode II.

It is important to note that we can investigate the ratio between
different networks such as $R_{{\rm E.}{\it coli}}(k_1,k_2)/R_{\rm
WWW}(k_1,k_2)$ because $P(k_1,k_2)$ shows a power law behavior.
Thus, even though the WWW and the metabolic network have different
ranges of the values of $k$, power law scaling of $P(k_1,k_2)$
implies that the ratio is independent on the region of $k_1$ and
$k_2$ used to plot this quantity.

\section{Invariance of degree distribution}
\label{invariant}

\begin{figure}
\centerline{ (a) \resizebox{7cm}{!} {
 \includegraphics{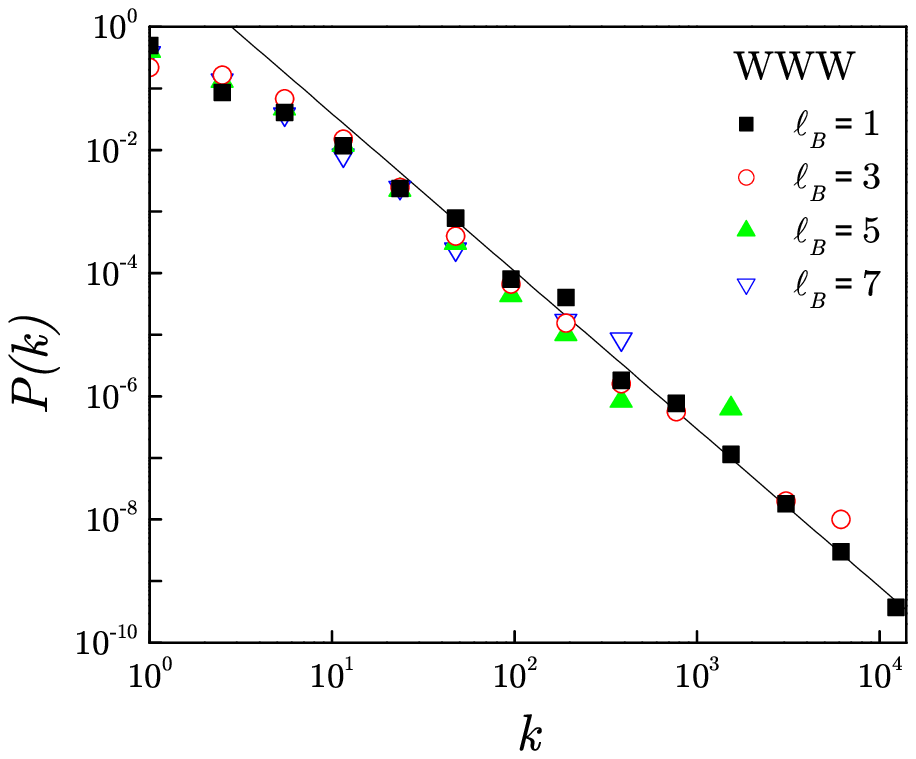}}
(b) \resizebox{7cm}{!} {
 \includegraphics{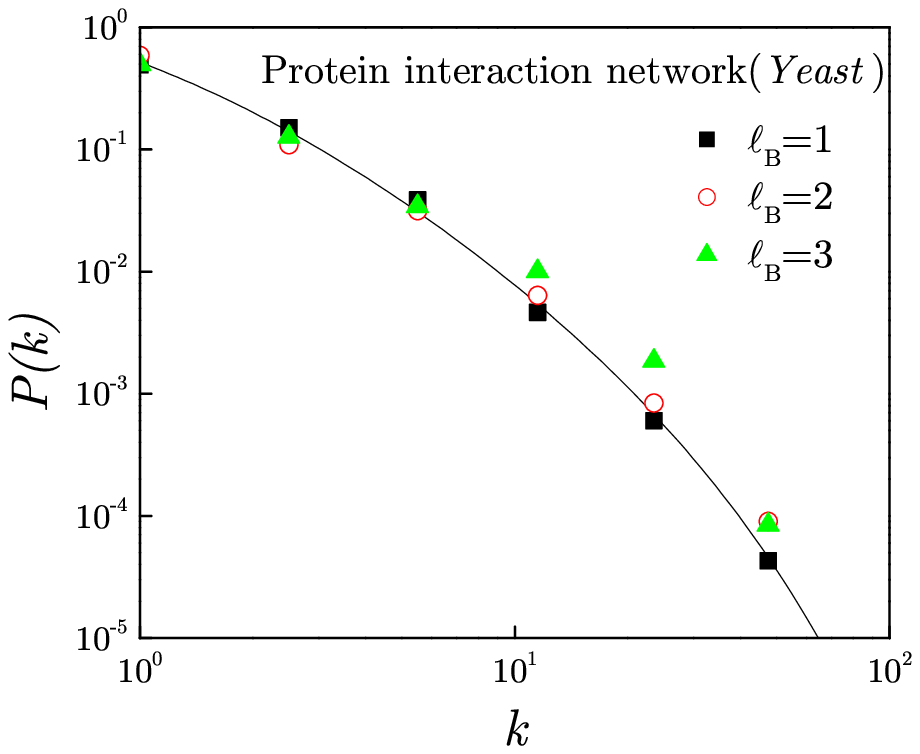}}
} \centerline{ (c) \resizebox{7cm}{!} {
\includegraphics{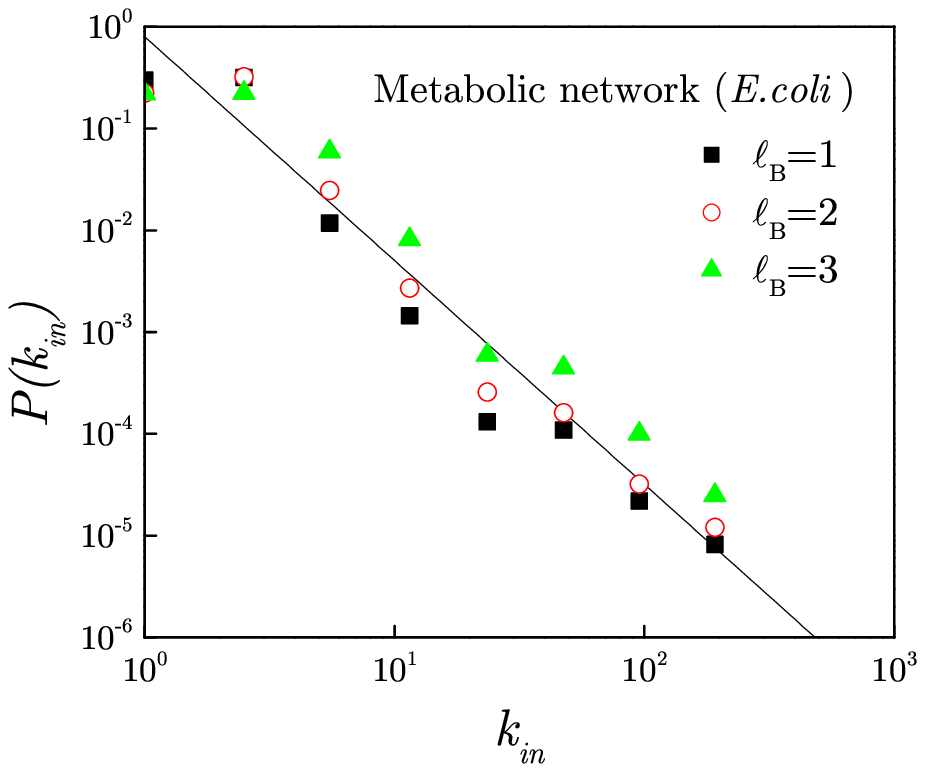}} (d)
\resizebox{7cm}{!} {
 \includegraphics{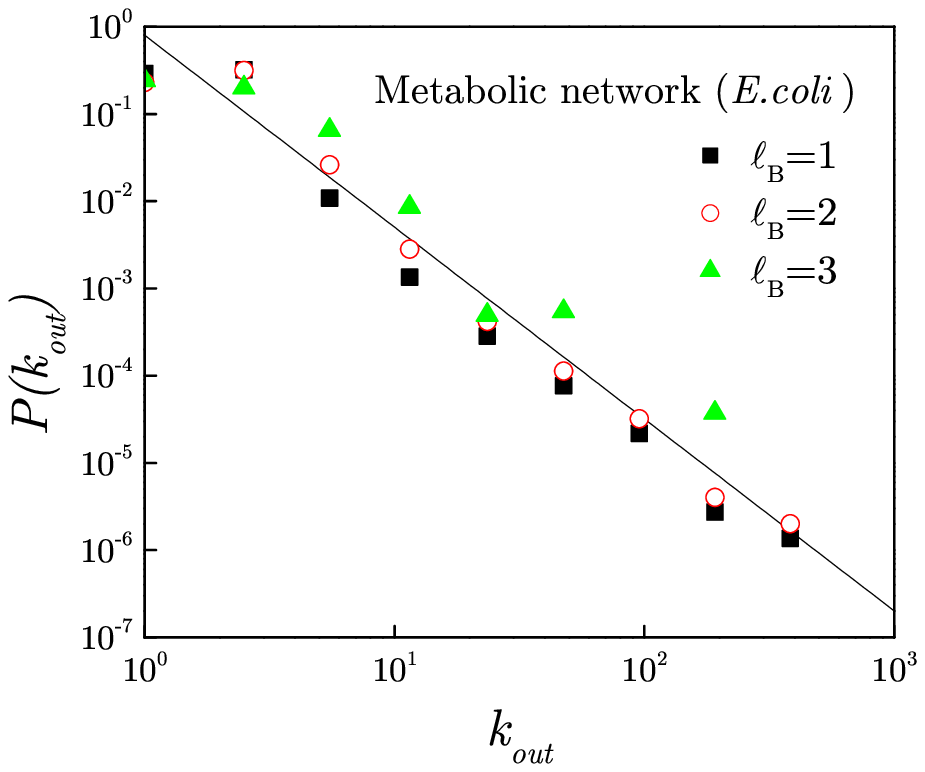}}
} \centerline{ (e) \resizebox{7cm}{!} {
\includegraphics{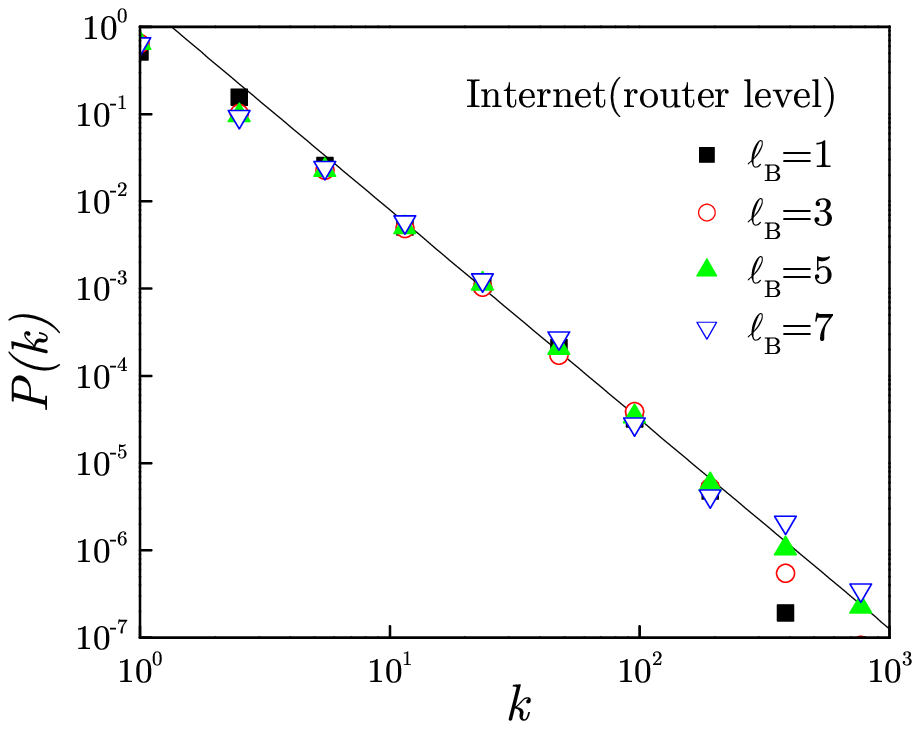}} }
\caption {Test of the invariance of degree distribution under
renormalization. We plot the degree distribution of (a) WWW
\cite{barabasi1999}, (b) Protein interaction network of {\it
yeast}, the inward degree (c) and the outward degree (d) of
metabolic network of E. {\it coli} \cite{cellular}, and (e)
Internet \cite{lucent}, according to different box size $\ell_B$.}
\label{pk_invariant}
\end{figure}

The renormalization procedure gives rise to a series of
coarse-grained networks based on the box length $\ell_B$. The
statistical properties of these networks, in particular, the
degree distribution keeps invariant, as we showed in the previous
work \cite{shm}. In this section, we verify this property again
for a wide range of real-world networks including both fractal
networks (WWW, protein interaction network of {\it yeast} and
metabolic network of E. {\it coli}) and non-fractal networks
(Internet), as we show in Fig. \ref{pk_invariant}. It is important
to note that even though the Internet is not fractal, the degree
distribution is still invariant under renormalization.

\section{Theory}
\label{theory}

Here we elaborate on several theoretical expressions presented in
the main text. We fully develop the theoretical framework of
renormalization and its analogy with the time evolution of
networks.

\begin{table}
\caption{ \label{table} Relation between time evolution and
renormalization.}
\begin{ruledtabular}
\begin{tabular}{lcccc}
Quantity
&Time evolution&Renormalization&\\
\hline \\
diameter&$\tilde L(t_1)+L_0$ & $L/(\ell_B+L_0)$& \\
&$\tilde L(t_2)+L_0$ & $L$ \\
\hline \\
number of &$\tilde N(t_1)$ & $N_B(\ell_B)$ &\\
nodes &$\tilde N(t_2)$ & $N$ \\
\hline \\
degree&$\tilde k(t_1)$ & $k(\ell_B)$& \\
&$\tilde k(t_2)$ & $k_{hub}$ \\
\hline \\
hub-hub&$\tilde k(t_1)$ & $k(\ell_B)$ &\\
links&$\tilde n_{h}(t_2|t_1)$ & $n_{h}(\ell_B)$ &\\
\end{tabular}
\end{ruledtabular}
\end{table}

The multiplicative growth law is expressed as:
\begin{equation}
\begin{split}
\tilde L(t+1) + L_0= a (\tilde L(t)+L_0),\\
\tilde N(t+1) = n \tilde N(t),\\\tilde k(t+1)=s \tilde k(t),\\
\tilde n_{h}(t+1) = e \tilde k(t),
\end{split}
\label{exp0}
\end{equation}
where all the quantities have been previously defined in the main
text. In Fig. \ref{renormalization}a of the main text we provide
an example of these quantities in a hypothetical growth process.
In this example $\tilde{N}(t)=16$ nodes are renormalized with
$N_B(\ell_B)=4$ boxes of size $\ell_B=3$ so that
  $\tilde{N}(t-1)=4$ nodes existed in the previous time step.
The box size is defined as the maximum chemical distance in the
box plus one. The chemical  distance is the number of links of the
minimum path between two nodes. The central box has a hub with
$k_{hub}=8$ links, then $\tilde{k}(t) = 8$. After renormalization,
 $k_B(\ell_B=3)=3$ for this central box,
so that $\tilde{k}(t-1) = 3$. Out these three links, two are via a
hub-hub connection (Mode I), thus $n_h(\ell_B) = 2$ and ${\cal
E}(\ell_B) = 2/3$, for this case.

We obtain the relation between the quantities at two times $t_2 >
t_1$ as
\begin{equation}
\begin{split}
\tilde L(t_2) + L_0= a^{t_2-t_1} (\tilde L(t_1)+L_0),\\
\tilde N(t_2) = n^{t_2-t_1} \tilde N(t_1),\\\tilde
k(t_2)=s^{t_2-t_1} \tilde
k(t_1),\\
\tilde n_{h}(t_2|t_1) = e^{t_2-t_1} \tilde k(t_1).
\end{split}
\label{exp1}
\end{equation}

Notice that the quantity $\tilde n_{h}(t_2|t_1)$ represents a
special case. This quantity indicates the number of links at time
$t_2$, which are connected to hubs generated before time $t_1$. To
avoid the confusion with the other quantities,  we introduce a new
notation $\tilde n_h(t_2|t_1)$ instead of $\tilde n_h(t_2)$ as
used for the other quantities in  Eq. (\ref{exp1}). We also notice
that the notation $\tilde n_h(t)$ in the main text Eq. (\ref{e})
is then interpreted as $\tilde n_h(t|t-1)$ for short. We then
obtain: $\tilde n_h(t_2|t_1) = e ~\tilde n_h(t_2-1|t_1) = \ldots =
e^{t_2-t_1} ~ n_h(t_1|t_1)=e^{t_2-t_1} ~k(t_1)$, where  we have
used that $n_h(t_1|t_1)=k(t_1)$.

The relationship between the quantities describing the time
evolution and the renormalization is shown in Table \ref{table}.
They are formalized as follows:

\begin{equation}
\begin{split}
\ell_B + L_0 = (\tilde L(t_2) + L_0)/(\tilde L(t_1)+L_0) = a^{t_2-t_1} \\
{\cal N} (\ell_B) \equiv N_B(\ell_B)/N = \tilde N(t_1)/\tilde
N(t_2) =
n^{t_1-t_2},\\
{\cal S} (\ell_B) \equiv k_B(\ell_B)/k_{hub} = \tilde
k_B(t_1)/\tilde
k_B(t_2) = s^{t_1-t_2},\\
{\cal E} (\ell_B) \equiv n_{h}(\ell_B)/k_B(\ell_B) = \tilde
n_{h}(t_2|t_1)/\tilde k(t_1) = e^{t_2-t_1}.
\end{split}
\label{exp3}
\end{equation}
Here we define the additional ratios, $\cal N$ and $\cal S$.
Replacing the time interval $t_2 - t_1$ by  $\ln (\ell_B + L_0) /
\ln a$, as obtained from the first equation in (\ref{exp3}), we
obtain:
\begin{equation}
\begin{split}
{\cal N} (\ell_B) = (\ell_B + L_0)^{-\ln n /\ln a},\\
{\cal S} (\ell_B) = (\ell_B + L_0)^{-\ln s /\ln a},\\
{\cal E} (\ell_B) = (\ell_B + L_0)^{-\ln (1/e) /\ln a},
\end{split}
\label{exp4}
\end{equation}
or
\begin{equation}
\begin{split}
{\cal N} (\ell_B) = (\ell_B + L_0)^{-d_B}, d_B \equiv \ln n /\ln a,\\
{\cal S} (\ell_B) = (\ell_B + L_0)^{-d_k}, d_k \equiv \ln s /\ln a,\\
{\cal E} (\ell_B) = (\ell_B + L_0)^{-d_e}, d_e \equiv \ln (1/e)
/\ln a,
\end{split}
\label{exp5}
\end{equation}
which correspond to the equations described in the main text.
Notice that we have considered $L_0=0$ in Eqs. (\ref{fractal}) for
simplicity. Equations (\ref{exp5}) are more general and
accommodate the case of non-fractal networks which are
characterized by exponential functions:

\begin{equation}
\begin{split}
{\cal N}(\ell_B) \sim\exp(-\ell_B/\ell_0),\\
{\cal S}(\ell_B) \sim \exp(-\ell_B/\ell'_0).
\end{split}
\label{exp6}
\end{equation}
These expressions arise from Eqs. (\ref{exp5}) by taking the limit
of $d_B\to \infty$, $d_k\to \infty$, and $L_0\to \infty$ while
$L_0/d_B \to \ell_0$ and $L_0/d_k \to \ell'_0$, where $\ell_0$ and
$\ell'_0$ are characteristic constants of the network.

\subsection{The minimal model}
\label{minimal}

\begin{figure}
\resizebox{10cm}{!} { \includegraphics{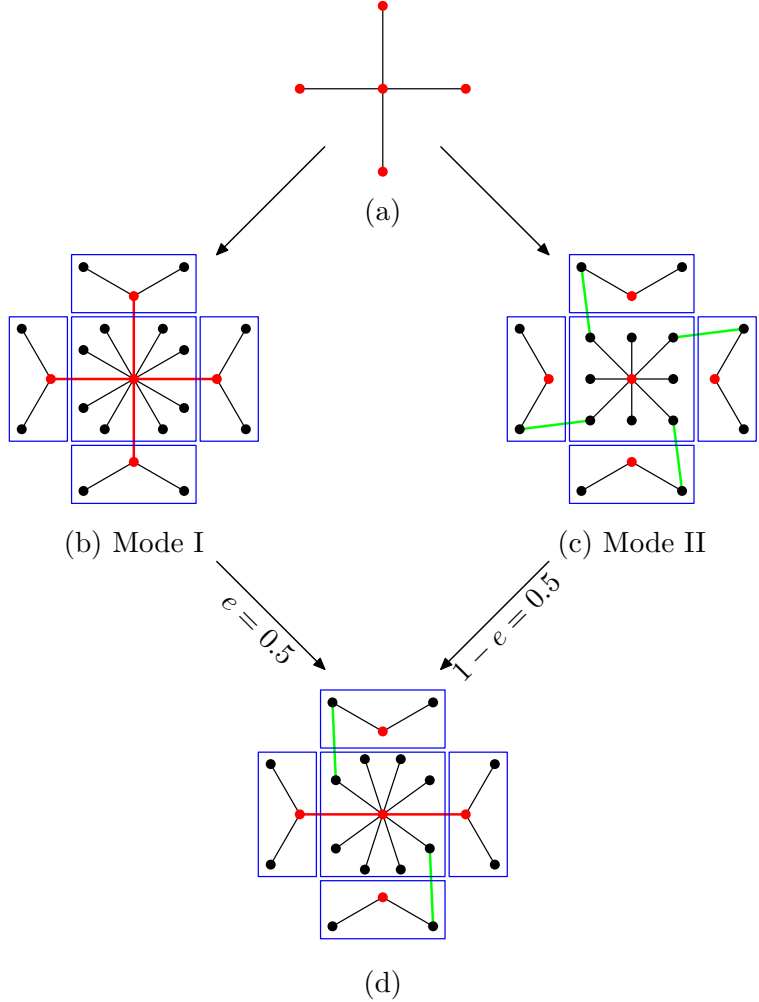}} \caption
{Different modes of growth with $m=2$. Starting with (a) five
nodes at $t=0$, the different connectivity modes lead to different
topological structures, which are (b) Mode I, (c) Mode II and (d)
combination of Mode I and II with probability $e=0.5$.
}\label{modes}
\end{figure}

\begin{figure}
\centerline{
 \resizebox{8cm}{!} {
 \includegraphics{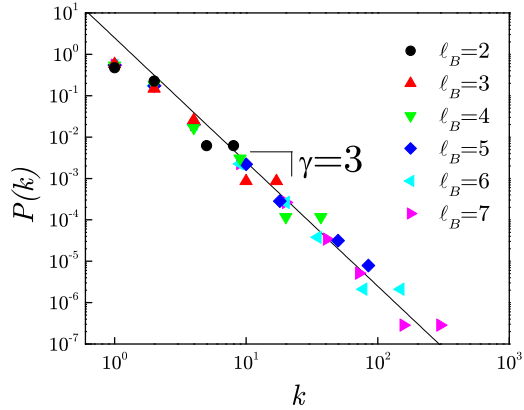}}
} \caption {Predictions of the model for $e=0.5$ for the
 degree distribution showing the power law behavior with
 $\gamma=3$ and its invariance under time evolution.
} \label{P(k)}
\end{figure}

In the framework of the minimal model, we start with a star
structure at $t=0$ as seen in Fig \ref{modes}a. At each time step
$m k(t)$ new nodes are generated for each node with degree $k(t)$,
where $m$ is an input parameter ($m=2$ in Fig. \ref{modes}).
Accordingly, we have $\tilde{N}(t+1)=\tilde{N}(t)+2m\tilde{K}(t)$,
where $\tilde{K}(t)$ is the total number of links at time $t$.
Since we do not consider the loop structure at the moment, we have
$\tilde{K}(t)=\tilde{N}(t)$. Then we obtain
$\tilde{N}(t+1)=(2m+1)\tilde{N}(t)$, or $n=2m+1$. We find that the
results of the model are independent on the initial configuration.

Then, two different connectivity modes are chosen as follows: Mode
I, we keep all the old connections generated multiplicatively at
time $t$ (the red links in Fig. \ref{modes}b). Mode II, all the
old connections generated in the previous time time step are
replaced by links between new generated nodes  (see the green
links in Fig. \ref{modes}c).

Mode I implies $s = m +1$, since $\tilde k(t+1) = m\tilde
k(t)+\tilde k(t)$, where the term $mk$ comes from newly generated
nodes, and the term $k$ comes from the links at previous time
steps. The diameter of the network grows additively as:
$\tilde{L}(t+1)=\tilde{L}(t)+2$ ($a=1$, $L_0\rightarrow\infty$ and
$(a-1)L_0\rightarrow 2$ in Eqs. (\ref{exp0})) because at each step
we generate one extra node at both sides of the network and
therefore the size of the network is increase by 2, as seen in
Fig. \ref{modes}b. This implies $\tilde{L}(t)\sim 2t$. For this
mode we obtain a non-fractal topology: $N_B(\ell_B)/N \sim
\exp(-\frac{\ln n}{2}\ell_B)$ and $k(\ell_B)/k_{hub}\sim
\exp(-\frac{\ln s}{2}\ell_B)$; a direct consequence of the linear
growth of the diameter $\tilde L(t)$ which implies that the
network is small-world. Moreover, $a=1$ leads to $d_B=\ln n/\ln a
\to \infty$ in this case (non-fractal).

Mode II alone gives rise to a fractal topology but with a
breakdown of the small-world property. The diameter increases
multiplicatively $\tilde{L}(t+1)=3\tilde{L}(t)$ ($a =3$ and
$L_0=0$ in Eqs. (\ref{exp0})), because we replace all the links at
previous time step by the paths with chemical distance $3$. The
degrees grow as $\tilde k(t+1)=m\tilde k(t)$ according to our
generation protocol, which leads to $s = m$. The multiplicative
nature of $\tilde{L}(t)$ leads to an exponential growth in the
diameter with time, $\tilde{L}(t)\sim e^{t\ln 3}$, and
consequently to a fractal topology with finite $d_B$ and $d_k$
according to Eqs. (\ref{exp3})-(\ref{exp4}). This is seen because
for this mode we have $a=3$. We obtain $N_B(\ell_B)\sim
\ell_B^{-d_B}$ with finite $d_B = \ln(2m+1)/\ln 3$ and also
$k(\ell_B)/k_{hub}\sim\ell_B^{-d_k}$ with $d_k = \ln m/\ln 3$.
However, the multiplicative growth of $\tilde{L}$ leads to the
disappearance of the small-world effect, which is replaced by a
power-law dependence.

The general growth process is a stochastic combination of Mode I
(with probability $e$) and Mode II (with probability $1-e$, see
Fig. \ref{modes}d). We obtain $\tilde{L}(t+1) =
(3-2e)\tilde{L}(t)+ 2e$, and $a = 3-2e$ and $L_0 = e/(1 - e)$.
Then, when $e\rightarrow 1$ (Mode I) $a = 1$ and $d_B
\rightarrow\infty$, $L_0\rightarrow\infty$ and we obtain a
non-fractal topology. On the other hand, Mode II has $e = 0$, then
$L_0 = 0$ and $a > 1$ and $d_B$ is finite, following the fractal
scaling. For an intermediate $ 0 < e < 1$ this model predicts
finite fractal exponents $d_B$ and $d_k$ and also predicts the
small-world effect due to the presence of Mode I, as shown in the
main text. This is seen in the exponential behavior of $\langle
M_c\rangle$ versus $\ell_B$.

In Fig. \ref{P(k)} we show further evidence that this model
reproduces the self-similar properties found in fractal networks
by plotting $P(k)$. We find that the model is modular since $P(k)$
is invariant under renormalization with $ \gamma = 1 + \ln n/ \ln
s = 3$, which is in agreement with the empirical findings of Fig.
\ref{pk_invariant}. Consistent with our predictions we find that
$d_k = \ln s/ \ln a = 3.3$.

Finally we summarize the predictions of the model. The fractal
dimension is $d_B = \ln n / \ln a$, and in the framework of the
minimal model with probability $e$, we find $a=3-2e$ and
$L_0=e/(1-e)$. Then, when $e\to 1$ (Mode I) $a=1$ and $d_B\to
\infty$, $L_0\to \infty$ and $\ell_0=L_0/d_B= 2/\ln n$ giving  a
non-fractal topology as in Eqs. (\ref{exp6}). On the other hand,
Mode II has $e=0$, then $L_0=0$ and $a > 1$ and $d_B$ is finite,
 following the fractal scaling of Eqs. (\ref{exp5}),
as long as the growth of the number of nodes is multiplicative
with a well-defined value of $n$. The fact that $a=1$ implies a
linear growth of the diameter $\tilde L(t) \sim 2 t$, which
produces the small-world property. An intermediate model with, for
instance $e=0.8$ gives rise to a fractal network with the small
world effect, as shown by the scaling of $N_B(\ell_B)$ in  Fig.
\ref{model}c and ${\cal E}(\ell_B)$ in Fig. \ref{model}d, in the
main text.

\subsection{Additional supporting evidence for the fractal
network model}
Evidence is given in Fig. \ref{model-mb}a for the
minimal model with parameter $e=0.8$. We calculate {\it (i)} the
mean number of nodes (mass) of the boxes tiling the network,
$\langle M_B (\ell_B)\rangle$ ($\equiv N / N_B(\ell_B)$), by using
the box covering methods, and {\it (ii)} the local mass $\langle
M_c(\ell_c)\rangle$ by averaging over  boxes of size $\ell_c$
around a randomly chosen node
 (the cluster growing method, see
 \cite{shm,vicsek}). The
results show how the minimal model reproduces one of the main
properties of fractal networks \cite{shm}: the power-law relation
for the global average mass $\langle M_B (\ell_B)\rangle \sim
\ell_B^{d_B}$ with $d_B=\ln 5/\ln 1.4 = 4.8$ as a signature of
fractality consistent with Eq. (\ref{fractal}), and the
exponential dependence of the local mass $\langle
M_c(\ell_c)\rangle\sim e^{\ell_c/\ell_0}$ as a signature of the
small-world effect: $\ell_c \sim \ln \langle M_c\rangle$. Note
that the cluster growing method is actually a way to measure the
distance, while the box covering method measures the fractality
\cite{shm}. The model leads also to a smooth monotonic scaling in
the size distribution of modules as observed in \cite{shm}. The
global small-world properties are treated next.

\section{Global small world: short cuts in the network}
\label{small-world}

\begin{figure}
\centerline{(a) {\resizebox{8cm}{!} {
\includegraphics{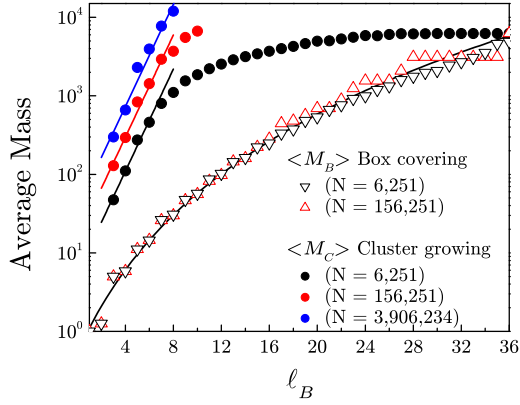}}} } \centerline{ (b)
\resizebox{8cm}{!} {
 \includegraphics{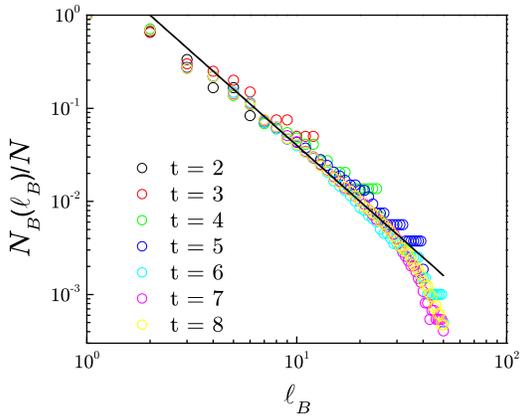}}
(c) \resizebox{8cm}{!} {
 \includegraphics{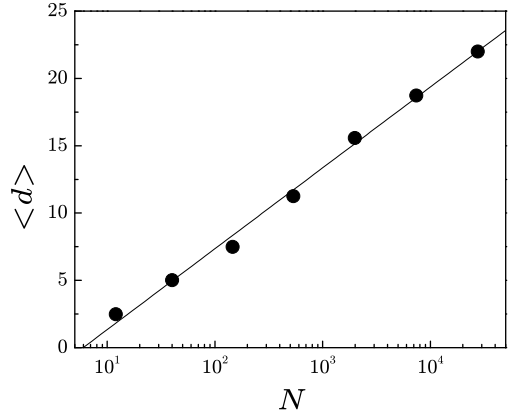}}
} \caption {Additional predictions of the renormalization growth
mechanism of complex networks. (a)
 The hallmark of fractality \cite{shm}
is predicted by the minimal mechanism: the global tiling of the
network evidences the fractality in the power-law dependence of
the mass of the boxes, $\langle M_B\rangle$, while the local
 average of the cluster growing method $\langle M_c\rangle$,
evidences the small-world effect. We show  simulations for $e=0.8$
and different network size. The scaling of
 $\langle M_B \rangle$ does not show finite
size effects. The initial exponential dependence range of
 $\langle M_c\rangle$ increases as the
size of the network increases. (b) $N_B$ vs $\ell_B$ for the model
with $e=0.5$ shows that the fractality still holds in the presence
of random noise. The
 straight line gives the theoretical prediction
of the model $d_B = 2$. (c) Average of the shortest path between
two nodes as a function of the system size showing the global
small world for the model ($e=0.5$). } \label{model-mb}
\end{figure}

An important factor in the dynamics of real-world networks is the
existence of randomness or noise in the growth process. The
simplest type of noise is the appearance of random connections
between nodes as exemplified in the Watts-Strogatz model of small
world networks \cite{watts}. To investigate how noise affects the
fractality of networks, we modify the dynamical law of the model
as follows: at each time step, $p \tilde K(t)$ number of links are
added in at random, here $\tilde K(t)$ is the total number of
links at time $t$, and $p$ is a constant that controls the
fraction of noise. We build a fractal, small-world and scale-free
topology with parameters $m = 1.5$, $e = 0.5$, and add $p = 1\%$
random connections at each time step. Our analytical
considerations predict a box dimension $d_B = 2$ in the absence of
noise. The numerical simulation (see Fig. \ref{model-mb}b) shows
that this prediction of $d_B$ still fits well to the simulated
data, except for a small deviation at large box sizes, i.e. the
added noise appears as an approximate exponential tail at large
distances. Interestingly, this method could be used to test the
appearance of noise in real complex networks, or to asses the
quality of the data in, for instance, protein interaction networks
obtained by yeast two-hybrid methods which are known to suffer
from many false positives.

Most interestingly, the addition of noise leads to the small-world
effect at the global level. In principle the existence of
fractality seems to be at odds with the small-world effect.
Fractality implies a power-law dependence on the distance, while
the small-world effect implies an exponential dependence
\cite{shm}. In Fig. \ref{model-mb}a we show how the combination of
Mode I and Mode II of growth leads to the global fractal property
and the local small-world effect. In Fig. \ref{model-mb}c we show
that by adding a small fraction of short-cuts in a fractal complex
network, we reproduce also the small-world effect at the global
level. Using the algorithm explained above we add noise to the
system and we find that the average distance $<d>$ over all pairs
of vertices is
\begin{equation}
<d> \sim 2.61\ln N,
\end{equation}
 (Fig. \ref{model-mb}c),
indicating that the fractal model also predicts the global
small-world. We notice that the fraction of short cuts needed to
obtain the global small-world is very  small, around $1\%$.

\section {Resilience of fractal networks under
intentional attack} \label{attack}

To compare the stability of fractal and non-fractal networks under
intentional attack (by removing hubs one by one from the largest
to the smallest one), we generate two networks with $e = 0$ and $e
= 1$. In general the threshold of collapse under attack
 depends on several parameters and not
only on the correlated properties of the network. Since we wish to
asses only the effects of anticorrelation for the vulnerability of
the network, we set all the other parameters to be equal. Thus, we
use the same $N$, $<k>$, $\gamma$, and also the same number of
loops in the structure. For this purpose we consider the number of
intraloops inside the boxes and the number of interloops between
boxes.

In practice, inside the box there are $mk(t)$ newly generated
nodes. We add  $y m k$ extra links between them to generate
triangles ($y$ is a given constant) to obtain loops inside the
boxes. For this case, we can rewrite $n = 2(1+y)m+1$, and the
clustering coefficient $C(k) = (2ym/s)k^{-1}$. Thus, this kind of
loops give rise to the known scaling of the clustering coefficient
with $k$ \cite{ravasz}.

Another type of loops appears when more than one link connects two
boxes (interloops). We find empirically that these kind of loops
are also arranged in a self-similar way and are characterized by a
new scaling exponent. In the framework of the minimal model, this
type of loops can be introduced by adding $x$ number of links
between boxes at each time step, instead of keeping one link
between boxes. These links could be of type Mode II (i.e., links
between non-hubs), or otherwise could be between a hub from one
box to a non-hub in the other node. In fact, this last mode of
growth is a third mode that can be considered in the minimal
model. We have not included it  so far for simplicity, since it
does not give rise to any new result. In general this mode could
be thought of as a modified Mode II, and does not change the
general conclusions of this study.

Combining the loop structure inside the boxes (intraloops
characterized by $y$) and between boxes (interloops characterized
by $x$) we obtain a general formula for the average degree
 $<k> = 2(1+y)+(x-1)/m$. In the case of the minimal tree
structure discussed in the main text we have $y = 0$ and $x = 1$,
which leads to $<k> = 2$, consistent  with our previous arguments.
These networks are then used to generate the structures used in
the calculation of the vulnerability of networks under intentional
attack, shown in Fig. \ref{modular}.

\end{document}